\documentclass[amsmath,floats,floatfix,11pt,nofootinbib,tightenlines,showpacs]{revtex4}
\usepackage{graphicx}
\usepackage{amsmath}
\usepackage{subfigure} 
\usepackage{amssymb}
\usepackage{bm}
\usepackage[usenames]{color}

\usepackage{ulem}
\newcommand{\shift}{\beta}
\newcommand{\Psileft}{\Psi^{L}}
\newcommand{\Psiright}{\Psi^{R}}
\DeclareMathOperator{\sgn}{sgn}
\newcommand{\ljump}{\big[\!\big[}
\newcommand{\rjump}{\big]\!\big]}
\newcommand{\bracketl}{\big\{\!\big\{}
\newcommand{\bracketr}{\big\}\!\big\}}
\begin{document}
\title{Discontinuous Galerkin method for computing gravitational\\
            waveforms from extreme mass ratio binaries}
      \author{
              Scott E.~Field${}^{1,}$\footnote{
              {\tt Scott\_Field@brown.edu},
              ${}^\dagger\,{}${\tt Jan\_Hesthaven@brown.edu},
              ${}^\ddagger\,{}${\tt srlau@math.unm.edu}},
              Jan~S.~Hesthaven${}^{2,\dagger}$, and
              Stephen R.~Lau${}^{2,3,\ddagger}$
              }
     \affiliation{
              ${}^1$Department of Physics, Brown
               University, Providence, RI 02912\\
              ${}^2$Division of Applied Mathematics, Brown
               University, Providence, RI 02912\\
              ${}^3$Mathematics and Statistics, University
               of New Mexico, Albuquerque, NM 87131
             }
\begin{abstract}
Gravitational wave emission from extreme mass ratio binaries
(EMRBs) should be detectable by the joint NASA--ESA LISA project, 
spurring interest in analytical and numerical methods for 
investigating EMRBs. We describe a  discontinuous Galerkin 
(dG) method for solving the distributionally forced 
1+1 wave equations which arise when modeling EMRBs via the 
perturbation theory of Schwarzschild blackholes. Despite the 
presence of jump discontinuities in the relevant polar and axial 
gravitational ``master functions", our dG method achieves global 
spectral accuracy, provided that we know the instantaneous 
position, velocity, and acceleration of the small particle.
Here these variables are known, since we assume that the 
particle follows a timelike geodesic of the Schwarzschild 
geometry. We document the results of several numerical 
experiments testing our method, and in our concluding section 
discuss the possible inclusion of gravitational self--force
effects. 
\end{abstract}
\maketitle
%
%
\section{Introduction}\label{sec:intro}

An extreme mass ratio binary (EMRB) is a system comprised of small 
mass--$m_p$ ``particle" (possibly a main sequence star, neutron star, 
or stellar mass blackhole) orbiting a large mass--$M$ blackhole, where 
the mass ratio $\mu = m_p/M \ll 1$. EMRB systems are expected to emit 
gravitational radiation in a low frequency band ($10^{-5}$ to $10^{-1}$ 
{\tt Hz}), and therefore offer the promise of detection by the 
joint NASA--ESA LISA project \cite{NASA_LISA,ScottHughes_LISA}. A 
standard method for studying some EMRBs uses the perturbation theory 
of Schwarzschild blackholes in an approximation which treats the particle 
as point--like and responsible for generating small metric perturbations
which radiate away to infinity. These perturbations 
influence the trajectory of particle, resulting in deviation from geodesic 
motion. Nevertheless, as a first and useful approximation, one may compute
the emitted gravitational radiation, assuming that the particle worldline 
is a timelike geodesic in the Schwarzschild spacetime. More sophisticated 
approaches have included the effect of radiation reaction on the particle, 
usually through incorporation of a self--force (see Ref.~\cite{PoissonLRR} 
for a review) or a suitable approximation thereof 
\cite{ScottHughes_Adiabatic}. Save for comments in the conclusion, 
this paper ignores self--force, although our methods may 
prove useful when it is included.

A chief goal of gravitational wave signal analysis is to determine the 
spacetime structure of a given binary system from its experimentally 
measured waveform \cite{ScottHughes_LISA,Barack_sources}, a goal likely 
facilitated by numerical simulation. For the scenario we consider,
simulation of ERMBs entails numerical evolution of Schwarzschild 
perturbations. The theory of such perturbations is well studied, and
starts with pioneering investigations by Regge and Wheeler \cite{RW}
and by Zerilli \cite{ZER}. A detailed presentation of the subject is 
beyond the scope of our paper, and we point the reader to 
Refs.~\cite{Martel_CovariantPert,SopuertaLaguna} for modern accounts
of the subject which supply the necessary background and 
references for our work. Nevertheless, in order to provide some context,
we now give a brief overview. We consider a small perturbation 
$\delta g_{\mu\nu}$ of the background Schwarzschild metric $g_{\mu\nu}$ 
given in \eqref{eq:staticelement}, where the perturbation satisfies the 
linearized Einstein equation. In our scenario 
the stress--energy tensor $T_{\mu\nu}$ given in \eqref{eq:IntTAB}
corresponds to a material point particle, and is therefore a distribution. 
The metric perturbation $\delta g_{\mu\nu}$ is clearly 
tensorial; nevertheless, the perturbations can
be reconstructed from a collection of scalar {\em master functions}. 
Remarkably, the master functions are governed by forced scalar wave 
equations with the following form:\footnote{
\label{fn:FandG} We could instead work with the equation
$$
-\partial_t^2 \Psi_{\ell m} + \partial_x^2 \Psi_{\ell m}
- V_\ell(r) \Psi_{\ell m} = 
\mathcal{G}_{\ell m}(t) \delta(r-r_p(t)) + 
\mathcal{F}_{\ell m}(t) 
\delta'(r-r_p(t)),
$$
where $\mathcal{G}_{\ell m}(t)$ and $\mathcal{F}_{\ell m}(t)$
depend only on $t$, and not on $r$. The relationships between
$\mathcal{G}_{\ell m}(t)$ and $\mathcal{F}_{\ell m}(t)$ and
our $G_{\ell m}(t,r)$ and $F_{\ell m}(t,r)$ follows from 
comparison between the right--hand sides of
Eq.~\eqref{eq:distforceRHS} and the last equation.}
\begin{equation}\label{eq:genericwaveeq}
-\partial_t^2 \Psi_{\ell m} + \partial_x^2 \Psi_{\ell m}
- V_\ell(r) \Psi_{\ell m} = f(r) \big[
G_{\ell m}(t,r) \delta(r-r_p(t)) + F_{\ell m}(t,r) \delta'(r-r_p(t))\big].
\end{equation}
The coordinates here are the areal radius $r$, the Regge--Wheeler tortoise 
coordinate \cite{MTW} $x = r + 2M\log(\frac{1}{2}r/M - 1)$, and the 
time--dependent radial location $r_p(t)$ of the particle. Furthermore, 
$V_\ell(r)$ is a potential, and $\Psi_{\ell m}(t,r)$ is one of the master 
functions. Since master functions arise in a (tensor) spherical harmonic 
decomposition of the perturbative equations for $\delta g_{\mu\nu}$, they 
carry multipole indices $(\ell,m)$, where $\ell \geq 2$, $|m| \leq \ell$. 
The {\em distributional} inhomogeneity on the right--hand side of 
\eqref{eq:genericwaveeq} stems from $T_{\mu\nu}$, and it involves Dirac delta 
functions, as well as the ordinary functions 
$F_{\ell m}(t, r)$, $G_{\ell m}(t, r)$, 
and $f(r) = 1 - 2M /r$. Perturbations of the Schwarzschild metric are
characterized as either {\em polar} or {\em axial}, and each case 
corresponds to particular functions. The polar case corresponds to the 
Zerilli potential \cite{ZER}
\begin{equation}\label{eq:ZerilliV}
V^\mathrm{Z}_\ell(r) = \frac{2f(r)}{(n_\ell r + 3 M)^2}
\left[n_\ell^2 \left(1 + n_\ell + \frac{3M}{r}\right) +
\frac{9M^2}{r^2} \left(n_\ell + \frac{M}{r}\right)\right],
\end{equation}
where $n_\ell = (\ell+2)(\ell-1)/2$, and the Zerilli--Moncrief master 
function $\Psi^\mathrm{ZM}$. Appendix \ref{sec:SourceDetails} provides
explicit formulas for $F^\mathrm{ZM}_{\ell m}(t, r)$ and 
$G^\mathrm{ZM}_{\ell m}(t,r)$. The axial case corresponds to the 
Regge--Wheeler potential \cite{RW}
\begin{equation}\label{eq:ReggeWheelerV}
V^\mathrm{RW}_\ell(r) = \frac{f(r)}{r^2}\left[\ell(\ell+1) 
- \frac{6M}{r}\right],
\end{equation}
and for this case we choose to work with the 
Cunningham--Price--Moncrief master function 
$\Psi^\mathrm{CPM}$ \cite{SopuertaLaguna}. Appendix 
\ref{sec:SourceDetails} also gives explicit formulas for 
$F^\mathrm{CPM}_{\ell m}(t, r)$ and 
$G^\mathrm{CPM}_{\ell m}(t,r)$.

A number of numerical methods for solving \eqref{eq:genericwaveeq} 
as an initial boundary value problem, and therefore modeling EMRBs
in the time--domain, have appeared in the literature. In particular, 
we note Lousto's fourth--order algorithm \cite{Lousto} based on
spacetime integration of \eqref{eq:genericwaveeq} and careful 
Taylor series arguments, and Sopuerta and Laguna's adaptive 
finite--element approach \cite{SopuertaLaguna}. Jung, Khanna, and 
Nagle have applied a spectral collocation method to the 
perturbation equations for head--on collisions, using spectral 
filtering to handle the delta function terms \cite{JungKhannaNagle}.
Most recently, Canizares and Sopuerta have proposed a multidomain 
spectral collocation method, with the particle location chosen 
between spectral elements \cite{CanizaresSopuerta}. Clearly, the 
key difficulty to overcome is the distributional forcing; however,
the problem should be amenable to a high--order accurate method, since 
---apart from possible transients--- the solutions we seek to 
compute are everywhere smooth, except for jump discontinuities 
at the particle location. As a suitable high--order scheme for 
solving \eqref{eq:genericwaveeq}, we propose a discontinuous 
Galerkin (dG) method, and our approach shares some similarities 
with Refs.~\cite{SopuertaLaguna,JungKhannaNagle,CanizaresSopuerta}, 
in particular we also ensure that the particle always lies at the 
interface between domain intervals. DG methods are widely used for
wave--dominated problems, such as 
electromagnetic scattering \cite{HesthavenWarburtonEM,HEST}, and
in nonlinear fluid dynamics \cite{HEST}.  
Our work is one of the first applications of dG methods 
to the modeling of gravitational waves (see also \cite{Zumbusch}), 
and the first dG computation of gravitational metric perturbations 
driven by a point--particle. Improving upon low--order methods, 
our method achieves global spectral accuracy (see also 
Refs.~\cite{JungKhannaNagle,CanizaresSopuerta}).

This paper is organized as follows. Section \ref{sec:preliminaries}
provides further background necessary to understand the physical model.
In particular, this section discusses the particle motion, the
resulting jump conditions in the master functions, and a coordinate
transformation adapted to the particle history. This background allows
us to rewrite \eqref{eq:genericwaveeq} as a first--order system which
features only {\em undifferentiated} delta-functions in the forcing.
Section \ref{sec:dgscheme} describes our dG scheme as applied to the
first--order system obtained in the previous section. Here we focus
not only on the local representation of the solutions,
but also on how delta function terms are incorporated into the
numerical flux function. Section \ref{sec:results} documents the 
results of several experiments testing our method, and it compares our
results with others found in the existing literature. In the concluding
section, we discuss, in particular, the possible inclusion of 
radiation reaction. Several appendices collect technical results
not given in the main part of the paper.

%
%
\section{Preliminaries}\label{sec:preliminaries}
Throughout, we use both $\partial$ 
and subscript notation for partial differentiation. For 
example, $\partial_r\Psi$ and $\Psi_r$ are the same. 
We use an over--dot to denote $\partial / \partial t$
differentiation, and sometimes a prime for differentiation
by argument. The labels ($\ell$,$m$,$\mathrm{CPM/ZM}$) are 
suppressed throughout.

\subsection{Particle motion}\label{subsec:particlemotion}
In standard coordinates, the
Schwarzschild line--element reads
\begin{equation}
\label{eq:staticelement}
ds^2 = -f dt^2 + f^{-1} dr^2 + r^2 (d\theta^2 + \sin^2\theta d\phi^2),
\end{equation}
where $t$ labels the preferred static--time slices and, as mentioned, $r$
is areal radius. Owing to the spherical symmetry of the line--element,
we may assume, without loss of generality, that the particle trajectory
$(r_p(t),\theta_p(t),\phi_p(t)) = (r_p(t),\pi/2,\phi_p(t))$ lies
in the equatorial plane \cite{Chan_MathBlack}. Introducing the 
{\em eccentricity} constant $e$, {\em semi--latus rectum} constant $p$, 
and the parameterization $r_p(t) = pM/(1+e\cos\chi(t))$, we obtain 
the particle trajectory $(r_p(t),\phi_p(t))$ by integration of the 
following system which describes timelike geodesic
motion: \cite{Chan_MathBlack,SopuertaLaguna,CUT,Martel_GravWave}
\begin{subequations} \label{INT_EOM}
\begin{align}
\frac{d\phi}{dt} & =\frac{(p-2-2e\text{cos}\chi)(1+e\text{cos}\chi)^2}
{Mp^{3/2}\big[(p-2)^2-4e^2\big]^{1/2}}\\
\frac{d\chi}{dt} & = \frac{(p-2-2e\text{cos}\chi)(1+e\text{cos}\chi)^2
\big[p-6-2e\text{cos}\chi\big]^{1/2}}{Mp^2 \big[(p-2)^2-4e^2\big]^{1/2}}.
\end{align}
\end{subequations}
We use $\chi(t)$ rather than $r_p(t)$, since the former increases 
monotonically through radial turning points.
In our scenario, integration of the system \eqref{INT_EOM} is independent 
of \eqref{eq:genericwaveeq}. Therefore, we may view the particle path, 
and so the right--hand side of \eqref{eq:genericwaveeq}, as predetermined. 
We shall be interested in the
parameter restriction $0\leq e <1$, for which the motion occurs between two
turning points and the orbit is bounded. The periastron and apastron occur 
respectively at $pM/(1+e)$ and $pM/(1-e)$, and for $e = 0$ the orbit is 
circular. Measured in coordinate time $t$, an eccentric orbit 
executes a radial period in time $T_r$ given by \cite{CUT}
\begin{align} \label{eq:Orbit_Period}
T_r & = C \int_{0}^{2\pi} d\chi
(1+e\text{cos}\chi)^{-2}\left[1-\frac{2(3+e\text{cos}\chi)}{p}\right]^{-1/2}
\left[1-\frac{2(1+e\text{cos}\chi)}{p}\right]^{-1}\\
C & = p^{3/2}M\left[\left(1 - \frac{2}{p}\right)^2-
\left(\frac{2 e}{p} \right)^2 \right]^{1/2}. 
\nonumber
\end{align}
When $e \neq 0$, we average physical quantities over 4 radial 
periods, as defined by Eq.~(\ref{eqnAveragedQuant}).

\subsection{Jump conditions}\label{subsec:masterJumps}
The forcing (\ref{eq:genericwaveeq}) induces jump conditions 
on the master function. Derived in Appendix \ref{AppJumps},
these are the following:
\begin{subequations} \label{eq:Alljumps}
\begin{align}
(f_p^2(t) - \dot{r}_p^2(t))\ljump\Psi\rjump & = f_p(t)
F(t,r_p(t))
\label{eq:jumpPsi}\\
2\dot{r}_p(t)\partial_t\ljump\Psi\rjump &
+ (\ddot{r}_p(t) - f_p(t) g_p(t) )\ljump\Psi\rjump
+ (f_p^2(t) - \dot{r}^2_p(t))\ljump\Psi_r\rjump
\nonumber \\
& = f_p(t) G(t,r_p(t))
- g_p(t)F(t,r_p(t))
- f_p(t)F_r(t,r_p(t)),
\label{eq:jumpPsitPsiPsir}
\end{align}
\end{subequations}
where the subscript $r$ in $F_r(t,r)$ denotes partial 
differentiation with respect to the second slot, and
\begin{equation}
f_p(t) = f(r_p(t)),\qquad g_p(t) = f'(r_p(t))
\neq \partial_t f_p(t)
\label{eq:pshorthands}
\end{equation}
are shorthands. In (\ref{eq:Alljumps}) our notation for a 
time--dependent jump is, for example,
\begin{equation}
\ljump\Psi\rjump(t) \equiv 
\lim_{\epsilon \rightarrow 0^+}
\big[
\Psi(t,r_p(t)+\epsilon)-
\Psi(t,r_p(t)-\epsilon)\big].
\label{eq:jumpdefinition}
\end{equation}
Defining the particle velocity as $v_p(t) = \dot{x}_p(t)
= \dot{r}_p(t)/f_p(t)$, we see that (\ref{eq:jumpPsi}) has
the form
\begin{equation}
f_p(t)(1-v^2_p(t))\ljump\Psi\rjump = F(t,r_p(t)),
\end{equation}
confirming that the jump $\ljump\Psi\rjump$ is well--defined for
a subluminal particle speed, $|v_p| < 1$. Therefore, we
may safely make the substitution
\begin{equation}
\ljump\Psi\rjump = 
\frac{f_p(t) F(t,r_p(t))}{f_p^2(t) - \dot{r}_p^2(t)}
\label{eq:jumpPsinew}
\end{equation}
in all formulas which follow.
Differentiation of (\ref{eq:jumpPsinew}) gives
\begin{align}
\partial_t \ljump\Psi\rjump & = \frac{2f_p(t)\dot{r}_p(t)F(t,r_p(t))
\left[\ddot{r}_p(t) - f_p(t) g_p(t)
\right]
}{(f_p^2(t) - \dot{r}_p^2(t))^2}
\nonumber \\
& + \frac{g_p(t) \dot{r}_p(t) F(t,r_p(t)) + f_p(t)F_t(t,r_p(t))
+ f_p(t)\dot{r}_p(t)F_r(t,r_p(t))}{f_p^2(t) - \dot{r}_p^2(t)}.
\label{eq:jumpPsit}
\end{align}
Finally, we may express (\ref{eq:jumpPsitPsiPsir}) as 
\begin{align}
\ljump\Psi_r\rjump & =
\big[
-2\dot{r}_p(t)\partial_t\ljump\Psi\rjump
-(\ddot{r}_p(t)-f_p(t) g_p(t) )\ljump\Psi\rjump
\nonumber \\
& 
+ f_p(t) G(t,r_p(t))
- g_p(t)F(t,r_p(t))
- f_p(t)F_r(t,r_p(t))\big]
\big/(f_p^2(t) - \dot{r}^2_p(t)),
\label{eq:jumpPsir}
\end{align}
with the understanding that here $\ljump\Psi\rjump$
and $\partial_t\ljump\Psi\rjump$ respectively stand for 
(\ref{eq:jumpPsinew}) and (\ref{eq:jumpPsit}). Again 
note that $f_p^2(t) - \dot{r}^2_p(t) > 1$ for a subluminal
particle speed, whence the jumps 
$\partial_t\ljump\Psi\rjump$ and $\ljump\Psi_r\rjump$ 
given by Eqs.~(\ref{eq:jumpPsit}) and (\ref{eq:jumpPsir}) 
are finite. The formulas
\begin{equation}\label{eq:charjumps}
\ljump \Psi_t\rjump = \partial_t \ljump\Psi\rjump
- \dot{r}_p(t)\ljump\Psi_r\rjump,\qquad
\ljump\Psi_x\rjump = f_p(t)\ljump\Psi_r\rjump
\end{equation}
prove useful later.

\subsection{Coordinate transformation adapted to particle history}
Now assume that $x \in [a,b]$ specifies the computational domain
and the time--dependent particle location $x_p = x_p(t)$ obeys 
$a < x_p(t) < b$, $\forall t$. We enact the coordinate transformation
\begin{align}
t & = \lambda \\
x & = a
+ \frac{x_p - a}{\xi_p -a}(\xi-a)
+ \frac{(b-x_p)(\xi_p-a)-(x_p-a)(b-\xi_p)}{(\xi_p-a)(b-\xi_p)(b-a)}
(\xi-a)(\xi-\xi_p),
\label{eq:xofxi}
\end{align}
with the understanding that $x_p = x_p(\lambda)$ is explicitly
time--dependent. The transformation obeys the following criteria:
(i) $t$ and $\lambda$ label the same time 
slices; (ii) $x(\lambda,\xi_p) = x_p(\lambda)$, with 
$\xi_p =$ constant and $a < \xi_p < b$;
(iii) $x(\lambda,a) = a$ and $x(\lambda,b) = b$. We further require (iv) 
that the transformation is invertible on $[a,b]$. This will only hold 
provided the point
\begin{equation}\label{eq:criticalxi}
\xi_\mathrm{critical} = \frac{(\xi_p + a)(\xi_p - x_p(\lambda)) 
                      +(x_p(\lambda)-a)(b-\xi_p)}{(\xi_p-x_p(\lambda))} 
\end{equation}
lies outside of the interval $[a,b]$. This is not a restriction of our 
method {\em per se}, and a coordinate transformation satisfying 
conditions (i) through (iv) may always be found. We have chosen to 
work with this one only for its simplicity.

Differentiations of (\ref{eq:xofxi}) yield
\begin{align}
\frac{\partial x}{\partial\lambda} & =
\frac{(\xi-a)(b-\xi) x_p'(\lambda)}{(\xi_p-a)(b-\xi_p)} 
\\
& \nonumber \\
\frac{\partial x}{\partial\xi} & =
\frac{(2\xi-\xi_p - a)(\xi_p-x_p(\lambda))
+(x_p(\lambda)-a)(b-\xi_p)}{(\xi_p-a)(b-\xi_p)}
\label{eq:explicit_dxdxi}
\\
& \nonumber \\
\frac{\partial^2 x}{\partial\xi^2} & =
\frac{2(\xi_p- x_p(\lambda))}{(\xi_p-a)(b-\xi_p)},
\end{align}
and these expressions appear in later formulas.

Under the coordinate transformation, the line--element 
(\ref{eq:staticelement})
acquires a shift vector,
\begin{equation}
ds^2 = -N^2 d\lambda^2 + L^2(d\xi + \shift^\xi d\lambda)^2 
+ r^2 (d\theta^2 + \sin^2\theta d\phi^2).
\end{equation}
Here $N = f^{1/2}$, $L = f^{1/2} \partial x/\partial \xi$, with 
$f$ understood as $f(r(x(\lambda,\xi)))$. The 
shift vector is 
\begin{equation}\label{eq:explicit_shiftvector}
\shift^\xi
= \frac{\partial x/\partial \lambda}{\partial x/\partial \xi}
= \frac{(\xi-a)(b-\xi) x_p'(\lambda)
}{(2\xi-\xi_p - a)(\xi_p-x_p(\lambda))
+(x_p(\lambda)-a)(b-\xi_p)},
\end{equation} 
and we will also need
\begin{equation}
\frac{\partial \shift^\xi}{\partial\xi} = 
\frac{(A\xi^2 + B \xi + C) x_p'(\lambda)}{
\big[(2\xi-\xi_p - a)(\xi_p-x_p(\lambda))
+(x_p(\lambda)-a)(b-\xi_p)\big]^2},
\end{equation}
where $A = 2(x_p(\lambda)-\xi_p)$, $B = 2 (a b + \xi_p^2 - (a + b) 
x_p(\lambda))$, and $C = (a^2 + b^2) x_p(\lambda)
- a (b - \xi_p)^2 - b (a^2 + \xi_p^2)$. 
The velocity variable $v = L \shift^\xi /N = \partial x/\partial \lambda$
obeys
\begin{equation}
v(\lambda,a) = 0,\qquad v(\lambda,\xi_p) 
= x'_p(\lambda),\qquad v(\lambda,b) = 0.
\end{equation}
Since $|x'_p(\lambda)| < 1$, we have $|v(\lambda,\xi)| < 1$
uniformly in $\xi$, assuming appropriately chosen $\xi_p$, 
$a$, and $b$. The vector field $\partial/\partial \lambda$ is not
the Killing direction, and it does not point orthogonal to the 
constant--$\lambda$ slices. To relate the $\partial/\partial \lambda$ 
direction to the unit--normal\footnote{Here we use coordinate--free
abstract notation for the {\em vector fields} $u$, $\bar{u}$, and $n$.} 
$u$ of the slicing, first consider 
$g_{\lambda\lambda} = 
-N^2 + L^2(\shift^\xi)^2 = -N^2(1-v^2) = - (N/\gamma)^2$,
where $\gamma = (1-v^2)^{-1/2}$ is the relativistic factor.
Therefore, the vector field 
\begin{equation}
\bar{u} = \gamma N^{-1} \partial/\partial \lambda
\end{equation}
is normalized, and one of its integral curves is the 
particle history. From standard formulas
\begin{equation}
N^{-1}(\partial/\partial \lambda
- \shift^\xi \partial/\partial \xi)
=  \gamma^{-1} \bar{u} - v L^{-1} \partial/\partial \xi,
\end{equation}
so that $\bar{u} = \gamma u + v\gamma n$. Here $g(n,n) =
L^{-2} g_{\xi\xi} = 1$, whence $n = L^{-1}
\partial/\partial \xi$ is a normalized spacelike
vector field. These formulas show that the spacetime
dependent parameter $v$ determines a local boost in the
tangent space of each spacetime point in the coordinate
domain. At $\xi = \xi_p$ this boost relates the slice normal
$u$ to the particle direction.

\subsection{Wave equation as a first--order system in $(\lambda,\xi)$}
Retaining the same letter $\Psi$ to denote the wave field 
$\Psi(\lambda,r(x(\lambda,\xi)))$ in the new coordinates,
we introduce the gradient $\Phi = \partial_\xi\Psi$. Notice that
$\partial_x \Psi = f\partial_r\Psi$ in the old system corresponds to 
$(\partial x/\partial\xi)^{-1}\Phi$ in the new system. The following 
first--order system in the $(\lambda,\xi)$ coordinates corresponds to the 
original second--order wave equation (\ref{eq:genericwaveeq}):
\begin{subequations}\label{eq:firstordersys}
\begin{align}
\partial_\lambda\Psi & = \shift^\xi \Phi - \Pi \\
& \nonumber \\
\partial_\lambda\Pi & = \shift^\xi \partial_\xi \Pi
-(\partial x/\partial\xi)^{-1}\partial_\xi [(\partial x/\partial\xi)^{-1}\Phi]
+ V(r) \Psi + J_1 \delta(\xi-\xi_p)
\\
& \nonumber \\
\partial_\lambda \Phi & = \partial_\xi (\shift^\xi \Phi)
- \partial_\xi \Pi  + J_2 \delta(\xi-\xi_p),
\end{align}
\end{subequations}
where Eq.~(\ref{eq:firstordersys}a) defines the variable $\Pi$.
The $\lambda$--dependent functions
\begin{equation}\label{eq:J1andJ2}
J_1 = -\shift^\xi \ljump\Pi\rjump  + (\partial x/\partial \xi)^{-2}
       \ljump\Phi\rjump,\qquad
J_2 = -\shift^\xi \ljump\Phi\rjump + \ljump\Pi\rjump.
\end{equation}
implement the jump conditions collected in Section 
\ref{subsec:masterJumps}, where in terms of \eqref{eq:charjumps}
\begin{equation}
\ljump\Pi\rjump = -\ljump\Psi_t\rjump,\qquad
\ljump\Phi\rjump = (\partial x/\partial \xi)\ljump\Psi_x\rjump.
\end{equation}
The jumps \eqref{eq:J1andJ2} can be recovered by
integrating (\ref{eq:firstordersys}) 
against a test function over the region
$(\xi_p-\epsilon,\xi_p+\epsilon)$, performing an integration by parts,
and taking the $\epsilon\rightarrow 0^+$ limit.
Smooth terms vanish in the limit. 
Thus, the system (\ref{eq:firstordersys}), with this choice of 
$J_1$ and $J_2$, is the first order form of \eqref{eq:genericwaveeq} by 
construction. 

%
%
\section{Discontinuous Galerkin method}\label{sec:dgscheme}

Following Ref.~\cite{HEST},
this section describes the nodal discontinuous Galerkin (dG)
method used to numerically solve \eqref{eq:firstordersys}.
Ultimately, we adopt a method--of--lines strategy, and here describe
the relevant semi--discrete scheme which arises upon spatial
approximation of \eqref{eq:firstordersys} by the dG method. In all
numerical experiments considered later, we have carried out the
temporal integration with an explicit fourth--order Runge--Kutta 
method, either one of low--storage \cite{RK} or the classical one. DG 
methods incorporate and build upon finite--element (FE), 
finite--volume (FV), and spectral methods, and in this section
the reader will recognize the features our dG approach shares with 
these more traditional methods. For example, on each subdomain 
our approach features a weak formulation of Legendre 
collocation, and our technique for coupling subdomains draws 
on FV methods.

\subsection{Local approximation of the system \eqref{eq:firstordersys}}

Our computational domain $\Omega$ is the closed 
$\xi$--interval $[a,b]$. We cover $\Omega$ with $K>1$ 
non--overlapping intervals $\mathsf{D}^k  = [a^k,b^k]$,
where $a = a^1$, $b = b^K$, and $b^{k-1} = a^k$ for 
$k = 2,\cdots,K$. We further assume that the 
particle location $\xi_p = b^{k_p} = a^{k_p+1}$ lies 
at the endpoint shared by $\mathsf{D}^{k_p}$ and
$\mathsf{D}^{k_p+1}$, with $1 \leq k_p < K$. On each interval 
$\mathsf{D}^k$, we approximate each component of the system 
vector $(\Psi,\Pi,\Phi)$ by a local interpolating polynomial of 
degree $N$. For example,
\begin{equation}\label{eq:Psiapprox}
\Psi^k_h(\lambda,\xi) = 
\sum_{j=0}^N \Psi(\lambda,\xi_j^k) \ell^k_j(\xi)
\end{equation}
approximates $\Psi$, where $\ell^k_j(\xi)$ is the $j$th 
Lagrange polynomial belonging to $\mathsf{D}^k$,
\begin{equation}
\ell^k_j(\xi) = \prod^{N}_{\stackrel{\scriptstyle i = 0}{\scriptstyle i\neq j}}
\frac{\xi - \xi^k_i}{\xi^k_j-\xi^k_i}.
\end{equation}
Evidently, the polynomial $\Psi^k_h$ interpolates $\Psi$ at 
the $\xi^k_j$. To define the nodes $\xi^k_j$, consider the mapping 
from the unit interval $[-1,1]$ to $\mathsf{D}^k$,
\begin{equation}
\xi^k(u) = a^k + {\textstyle \frac{1}{2}}(1+u)(b^k - a^k),
\end{equation}
and the $N$+1 Legendre--Gauss--Lobatto (LGL) nodes $u_j$. The 
$u_j$ are the roots of the equation
\begin{equation}
(1-u^2)P_N'(u) = 0,
\end{equation}
where $P_N(u)$ is the $N$th degree Legendre polynomial, and the 
physical nodes are simply $\xi^k_j = \xi^k(u_j)$.
In vector notation the approximation \eqref{eq:Psiapprox} 
takes the form
\begin{equation}
\Psi^k_h(\lambda,\xi) = 
\boldsymbol{\Psi}^k_h(\lambda)^T 
\boldsymbol{\ell}^k(\xi),
\end{equation}
in terms of the column vectors
\begin{equation}
\boldsymbol{\Psi}^k_h(\lambda) 
= \big[\Psi(\lambda,\xi^k_0),\cdots,
\Psi(\lambda,\xi^k_N)\big]^T,\qquad
\boldsymbol{\ell}^k(\xi)
= \big[\ell^k_0(\xi),\cdots,\ell^k_N(\xi)\big]^T.
\end{equation}
We also need to approximate by polynomials various 
products, for example $(\partial x/\partial\xi)\Phi$. Such 
approximations are achieved through pointwise representations such as
\begin{equation}\label{eq:productexpansion}
(x_\xi\Phi)^k_h(\lambda,\xi) = 
\sum_{j=0}^N x_\xi(\lambda,\xi^k_j)
\Phi(\lambda,\xi^k_j)\ell^k_j(\xi).
\end{equation}
Here, and in what follows, we use the shorthands 
$x_\xi = \partial x/\partial\xi$ and 
$x_{\xi\xi} = \partial^2 x/\partial\xi^2$. Our vector notation
for this example will be $(x_\xi\Phi)^k_h(\lambda,\xi) =
(\boldsymbol{x}_\xi\boldsymbol{\Phi})^k_h(\lambda)^T
\boldsymbol{\ell}^k(\xi)$. 

On each interval $\mathsf{D}^k$ and for each solution component, 
we define local residuals,
\begin{subequations}\label{eq:residuals}
\begin{align}
(R_{\Psi})^k_h & = \partial_\lambda \Psi^k_h
                 - (\shift^\xi\Phi)^k_h + \Pi^k_h 
\\
(R_\Pi)^k_h & = \partial_\lambda \Pi^k_h
- \partial_\xi (\shift^\xi \Pi)^k_h
+ (\Pi \partial_\xi \shift^\xi)^k_h
+\partial_\xi (x_\xi^{-2}\Phi)^k_h
+(x_\xi^{-3} x_{\xi\xi} \Phi)^k_h - (V\Psi)^k_h
\\
(R_{\Phi})^k_h & = \partial_\lambda \Phi^k_h
- \partial_\xi (\shift^\xi \Phi)^k_h 
+ \partial_\xi\Pi^k_h,
\end{align}
\end{subequations}
measuring the extent to which our approximations satisfy
the original system of PDE. We define these residuals on 
open intervals $(a^k,b^k) \subset \mathsf{D}^k$, but have 
assumed that the particle location $\xi_p = b^{k_p} 
= a^{k_p+1}$ lies at an endpoint. Therefore, in the 
residuals \eqref{eq:residuals} we have not yet included 
the $\delta$--function contributions appearing in 
\eqref{eq:firstordersys}.

To motivate our derivation of a numerical approximation to
\eqref{eq:firstordersys}, we first consider the $k$th inner 
product
\begin{equation}
(u,v)_{\mathsf{D}^k} \equiv \int^{b^k}_{a^k}d\xi u(\xi) v(\xi)
\end{equation}
and the expressions $(\ell^k_j,(R_{\Psi})^k_h)_{\mathsf{D}^k}$,
$(\ell^k_j,(R_{\Pi})^k_h)_{\mathsf{D}^k}$, and
$(\ell^k_j,(R_{\Phi})^k_h)_{\mathsf{D}^k}$. Namely, the 
inner products between the residual components 
\eqref{eq:residuals} and the $j$th Lagrange polynomial on 
$\mathsf{D}^k$. We call the requirement that {\em all} these
inner products $(\ell^k_j,(R_{\Psi})^k_h)_{\mathsf{D}^k}$, 
$(\ell^k_j,(R_{\Pi})^k_h)_{\mathsf{D}^k}$, 
$(\ell^k_j,(R_{\Phi})^k_h)_{\mathsf{D}^k}$
vanish $\forall j$ the $k$th {\em Galerkin conditions}.
For now, we focus on the $\Phi$ equation as a
representative example, but will later also consider the 
$\Psi$ and $\Pi$ equations. 

Enforcement of the Galerkin conditions on each 
$\mathsf{D}^k$ will not recover a meaningful global 
solution, since they provide no mechanism for coupling of
the individual local solutions on the different intervals. 
Notice that, upon integration by parts, 
we may express the inner product as follows:
\begin{align}\label{eq:innerlRPhi}
(\ell^k_j,(R_{\Phi})^k_h)_{\mathsf{D}^k} 
= & \int_{a^k}^{b^k}d\xi\big[\ell^k_j(\xi)\
  \partial_\lambda \Phi^k_h(\lambda,\xi)
+ \ell^k_j{}'(\xi)(\shift^\xi \Phi)^k_h(\lambda,\xi)
- \ell^k_j{}'(\xi)\Pi^k_h(\lambda,\xi)\big]
\nonumber \\
- & \left.\big[(\shift^\xi \Phi)^k_h(\lambda,\xi)
               -\Pi^k_h(\lambda,\xi)\big]
           \ell^k_j(\xi)\right|^{b^k}_{a^k}.
\end{align}
Therefore, in lieu of \eqref{eq:innerlRPhi} with
$(\ell^k_j,(R_{\Phi})^k_h)_{\mathsf{D}^k} = 0$, we 
enforce the equation
\begin{align}
0 = \int_{\mathsf{D}^k}d\xi\big[\ell^k_j(\xi) &
  \partial_\lambda \Phi^k_h(\lambda,\xi)
+ \ell^k_j{}'(\xi)(\shift^\xi \Phi)^k_h(\lambda,\xi)
- \ell^k_j{}'(\xi)\Pi^k_h(\lambda,\xi)\big]
\nonumber \\
- & \left.\big[\shift^\xi(\lambda,\xi)(\Phi^{k}_h)^* -
     (\Pi^{k}_h)^*\big]\ell^k_j(\xi)\right|^{b^k}_{a^k}.
\label{eq:innerlRPhiNum}
\end{align}
This equation features {\em numerical fluxes}, 
$-\shift^\xi(\lambda,a^k)(\Phi^k_h)^* +(\Pi^k_h)^*$ and 
$-\shift^\xi(\lambda,b^k)(\Phi^k_h)^* +(\Pi^k_h)^*$,
rather than boundary fluxes, 
$-(\shift^\xi \Phi)^k_h(\lambda,a^k) +\Pi^k_h(\lambda,a^k)$ and
$-(\shift^\xi \Phi)^k_h(\lambda,b^k) +\Pi^k_h(\lambda,b^k)$,
thereby coupling adjacent subdomains.
The numerical fluxes are determined by (as yet not chosen)
functions
\begin{equation}
(\Pi^k_h)^* = (\Pi^k_h)^*(\Pi^+_h,\Phi^+_h,\Pi^-_h,\Phi^-_h),\qquad
(\Phi^k_h)^* = (\Phi^k_h)^*(\Pi^+_h,\Phi^+_h,\Pi^-_h,\Phi^-_h),
\label{eq:PiPhifluxes}
\end{equation}
where, for example, $\Pi^-_h$ is an interior boundary value [either 
$\Pi^k_h(\lambda,a^k)$ or $\Pi^k_h(\lambda,b^k)]$ 
of the approximation defined on $\mathsf{D}^k$, and 
$\Pi^+_h$ is an exterior boundary value [either $\Pi^{k-1}_h(\lambda,b^{k-1})$ 
or $\Pi^{k+1}_h(\lambda,a^{k+1})$] of the approximation 
defined on either $\mathsf{D}^{k-1}$ or $\mathsf{D}^{k+1}$. The
fluxes (\ref{eq:PiPhifluxes}) could also depend on 
$\Psi^\pm_h$, but we will not need this extra generality.

Let us now write the $N$+1 equations \eqref{eq:innerlRPhiNum} in 
matrix form, and also collect the corresponding matrix forms 
associated with approximation of the $\Psi$ and $\Pi$ equations. 
To write down these matrix forms, we first introduce the $k$th
{\em mass} and {\em stiffness} matrices,
\begin{equation}
M_{ij}^k=\int_{a^k}^{b^k}d\xi \ell_i^k(\xi) \ell_j^k (\xi),\qquad
S_{ij}^k =\int_{a^k}^{b^k}d\xi \ell_i^k(\xi)\ell_j^k{}'(\xi).
\end{equation}
These matrices belong to $\mathsf{D}^k$, and the corresponding
matrices belonging to the reference interval $[-1,1]$ are 
\begin{equation}
M_{ij} = \int_{-1}^1 du \ell_i(u) \ell_j(u),\qquad
S_{ij} = \int_{-1}^1 du \ell_i(u) \ell_j'(u),
\end{equation}
where $\ell_j(u)$ is the $j$th Lagrange polynomial determined
by the LGL nodes $u_j$ on $[-1,1]$. These matrices are related 
by $M^k_{ij} = {\textstyle \frac{1}{2}}(b^k-a^k) M_{ij}$ 
and $S^k_{ij} = S_{ij}$, whence only the reference matrices
require computation and storage. In \eqref{eq:innerlRPhiNum}
we now expand all polynomial approximations in the Lagrange
polynomial basis as in \eqref{eq:Psiapprox}, thereby obtaining
\begin{equation}\label{eq:MSPhi}
M^k \partial_\lambda \boldsymbol{\Phi}^k_h + (S^k)^T
(\boldsymbol{\shift}^\xi\boldsymbol{\Phi})^k_h
- (S^k)^T \boldsymbol{\Pi}^k_h
= \left.\big[\shift^\xi(\lambda,\xi)(\Phi^{k}_h)^* -
  (\Pi^{k}_h)^*\big]\boldsymbol{\ell}^k(\xi)\right|^{b^k}_{a^k}.
\end{equation}
Now, as described in \cite{HEST}, the spectral collocation derivative 
matrix is 
\begin{equation}
(D^k)_{ij} = \left.\frac{d\ell^k_j}{d\xi}\right|_{\xi = \xi^k_i}.
\end{equation}
also given by $D^k = (M^k)^{-1} S^k$. By the symmetry of the mass 
matrix, $(D^k)^T = (S^k)^T (M^k)^{-1}$. We then introduce the 
similarity--transformed matrix
\begin{equation}
D_M^k = M^k D^k (M^k)^{-1},
\end{equation}
tailored to obey $(D_M^k)^T = (M^k)^{-1} (S^k)^T$.
Applying $(M^k)^{-1}$ to both sides of \eqref{eq:MSPhi} yields
an equation for $\partial_\lambda \boldsymbol{\Phi}^k_h$.
The result and the corresponding equations for 
$\partial_\lambda \boldsymbol{\Psi}^k_h$ and 
$\partial_\lambda \boldsymbol{\Pi}^k_h$ (derived via 
calculations similar to those presented above) are the 
following:
\begin{subequations}\label{eq:vectorsystem}
\begin{align}
& \partial_\lambda \boldsymbol{\Psi}^k_h - 
(\boldsymbol{\shift}^\xi\boldsymbol{\Phi})^k_h
+ \boldsymbol{\Pi}^k_h
= 0 \\
& \partial_\lambda \boldsymbol{\Pi}^k_h
+ (D^k_M)^T (\boldsymbol{\shift}^\xi\boldsymbol{\Pi})^k_h
- (D^k_M)^T (\boldsymbol{x}_\xi^{-2}\boldsymbol{\Phi})^k_h
+ (\boldsymbol{x}_\xi^{-3}\boldsymbol{x}_{\xi\xi}
   \boldsymbol{\Phi})^k_h
- (\boldsymbol{V}\boldsymbol{\Psi})^k_h
\nonumber \\
& 
\hspace{2cm}  = 
\left.(M^k)^{-1}\big[\left(\shift^\xi(\lambda,\xi)(\Pi^{k}_h)^* -
x_\xi^{-2}(\lambda,\xi)(\Phi^{k}_h)^*\right)\boldsymbol{\ell}^k(\xi)
\big]\right|^{b^k}_{a^k}
\\
& \partial_\lambda \boldsymbol{\Phi}^k_h + (D^k_M)^T
(\boldsymbol{\shift}^\xi\boldsymbol{\Phi})^k_h
- (D^k_M)^T \boldsymbol{\Pi}^k_h
= \left.(M^k)^{-1}\big[\left(\shift^\xi(\lambda,\xi)(\Phi^{k}_h)^* -
     (\Pi^{k}_h)^*\right)\boldsymbol{\ell}^k(\xi)\big] 
    \right|^{b^k}_{a^k}.
\end{align}
\end{subequations}
All adjacent vectors in these expressions, 
e.~g.~$(\boldsymbol{\shift}^\xi\boldsymbol{\Phi})^k_h$,
$(\boldsymbol{V}\boldsymbol{\Psi})^k_h$, and 
$(\boldsymbol{x}_\xi^{-3}\boldsymbol{x}_{\xi\xi}
\boldsymbol{\Phi})^k_h$, should be interpreted as a 
single vector obtained via component--by--component 
products. 

\subsection{Numerical Flux}

To define the vector $(f_\Pi,f_\Phi)^T$ of physical fluxes, we write 
(\ref{eq:firstordersys}b,c) as
\begin{equation}
\partial_\lambda 
\left(
\begin{array}{c}
\Pi 
\\
\Phi
\end{array}
\right)
+
\partial_\xi 
\left(
\begin{array}{c}
f_\Pi 
\\
f_\Phi
\end{array}
\right) = \text{lower order terms}.
\end{equation}
This equation determines the physical and numerical fluxes as 
follows:
\begin{equation}\label{eq: NF_matrix}
\left(
\begin{array}{c} 
f_{\Pi} \\ f_{\Phi} 
\end{array}
\right) 
\equiv
\left( 
\begin{array}{cc} 
-\shift^\xi & x_\xi^{-2} \\ 
 1          & -\shift^\xi 
\end{array}
\right)
\left(
\begin{array}{c} 
\Pi \\ \Phi 
\end{array}
\right),
\qquad
\left(\begin{array}{c}
(f_{\Pi}^k)^* \\ 
(f_{\Phi}^k)^* 
\end{array}\right) 
\equiv
\left(\begin{array}{cc} 
-\shift^\xi & x_\xi^{-2} \\ 
1           & -\shift^\xi 
\end{array}\right)
\left(\begin{array}{c}(\Pi^{k}_h)^* \\ 
(\Phi^{k}_h)^* \end{array}\right).
\end{equation}
The combinations of $(\Pi^{k}_h)^*$ and $(\Phi^{k}_h)^*$
which appear in (\ref{eq:vectorsystem}b,c) are precisely
$-(f_{\Pi}^k)^*$ and $-(f_{\Phi}^k)^*$, as must be the case
since these terms have arisen through integration by parts. 
In this subsection we construct the required boundary expressions 
for $(f_{\Pi}^k)^*$ and $(f_{\Phi}^k)^*$.
Our numerical flux must be robust, ensure stability, and 
be capable of handling the analytic discontinuities at the 
particle location. Numerical experiments suggest 
that inclusion of a Dirac delta function 
renders inadequate otherwise suitable numerical 
fluxes, such as the central and Lax--Friedrichs fluxes. 
However, as we will see, a suitably 
modified upwind numerical flux successfully handles the delta 
functions in the system \eqref{eq:firstordersys}, recovering 
optimal convergence. We begin by constructing the standard 
upwind flux corresponding to no particle, and then incorporate 
the particle's effect into the flux through the addition 
of an extra term.

An upwind numerical flux passes information across an interface 
in the direction of propagation. To construct the upwind numerical 
fluxes, we first diagonalize the matrix appearing in 
\eqref{eq: NF_matrix} as follows:
\begin{equation}\label{eq:Udiagonalize}
\left( \begin{array}{cc} 
-\shift^\xi & x_\xi^{-2} \\ 
1           & -\shift^\xi 
\end{array}\right) =
T \left( 
\begin{array} {cc}
-\shift^\xi + x_\xi^{-1} & 0 \\
0 & -\shift^\xi - x_\xi^{-1} 
\end{array}\right) T^{-1},
\qquad
T^{-1} = \left(\begin{array}{cc}
          1 & x_\xi^{-1} \\
          1 & - x_\xi^{-1}
          \end{array}\right).
\end{equation}
Application of $T^{-1}$ on the system vector $(\Pi,\Phi)^T$ 
of fundamental fields yields the system vector $(\Pi + \Phi/x_\xi,\Pi - 
\Phi/x_\xi)^T$ of characteristic fields. For our problem, the first 
characteristic field $\Pi + \Phi/x_\xi$ propagates rightward with 
speed $-\shift^\xi + x_\xi^{-1}$ relative to the $\partial/\partial\lambda$ 
time axis, while the second characteristic field $\Pi - \Phi/x_\xi$ 
propagates leftward with speed $-\shift^\xi - x_\xi^{-1}$. Respectively,
the upwind fluxes at a left endpoint $a^k$ ($k\neq k_p+1$) and at a
right endpoint $b^k$ ($k\neq k_p$) then take the following forms:
\begin{subequations}\label{leftright_upwind_form}
\begin{align}
\left(\begin{array}{c}(f_{\Pi}^k)^*
\\ (f_{\Phi}^k)^* \end{array}\right)_\mathrm{left} & =
T
\left( \begin{array}{cc}
0&0\\0& -\shift^\xi - x_\xi^{-1}
\end{array}\right)
T^{-1}
\left(\begin{array}{c}
\Pi_h^{-} \\ \Phi_h^{-}
\end{array}\right) +
T
\left( \begin{array}{cc}
-\shift^\xi + x_\xi^{-1}&0\\0&0
\end{array}\right)
T^{-1}
\left(\begin{array}{c}
\Pi_h^{+} \\ \Phi_h^{+}
\end{array}\right)
\\
\left(\begin{array}{c}(f_{\Pi}^k)^*
\\ (f_{\Phi}^k)^* \end{array}\right)_\mathrm{right} & =
T
\left( \begin{array}{cc}
0 & 0\\0& -\shift^\xi - x_\xi^{-1}
\end{array}\right)
T^{-1}
\left(\begin{array}{c}
\Pi_h^{+} \\ \Phi_h^{+}
\end{array}\right) +
T
\left( \begin{array}{cc} - \shift^\xi + x_\xi^{-1}&0\\0&0
\end{array}\right)
T^{-1}
\left(\begin{array}{c}
\Pi_h^{-} \\ \Phi_h^{-}
\end{array}\right).
\end{align}
\end{subequations}
Eqs.~(\ref{leftright_upwind_form}a,b)
formalize the intuitive concept behind the upwind numerical flux.
In these equations triple--product matrices operate on the interior and 
exterior solution. The first matrix operation
transforms the fields to characteristic fields, the second
projects out one of the characteristic fields, and the third
transforms back to the fundamental fields. As a result, 
information from a right--moving field, say, influences the 
subdomain to the right, but not the subdomain to the left.

To achieve succinct expressions for the upwind flux which hold at both
left and right endpoints, at each interface we define the average of a 
numerical variable and its numerical jump as
\begin{equation}
\bracketl \Phi \bracketr = \frac{1}{2}(\Phi^+ + \Phi^-),
\qquad
\ljump \Phi \rjump_\mathrm{num} 
= \mathbf{n}^+ \Phi^+ + \mathbf{n}^- \Phi^- .
\end{equation}
Here $\mathbf{n}$ denotes the local outward--pointing normal of a 
subdomain and can be $\pm 1$. The numerical jump here is not 
a predetermined analytical jump as defined in Eq.~\eqref{eq:jumpdefinition},
and it has a different sign convention. These definitions yield the following 
concise formulas (valid at left or right endpoints):
\begin{subequations}\label{eq:NumericalFlux}
\begin{align} 
(f_{\Pi}^k)^* & = 
\bracketl {-\shift^\xi} \Pi_h + x_\xi^{-2}\Phi_h \bracketr 
+ \frac{1}{2}
\ljump x_\xi^{-1}\Pi_h -x_\xi^{-1}\shift^\xi\Phi_h \rjump_\mathrm{num} 
\\
(f_{\Phi}^k)^* & = 
\bracketl \Pi_h  - \shift^\xi\Phi_h \bracketr 
+ \frac{1}{2}
\ljump x_\xi^{-1}\Phi_h -x_\xi\shift^\xi\Pi_h \rjump_\mathrm{num}.
\end{align}
\end{subequations}
At all interior endpoints ($a^k$ for $k\neq 1,k_p + 1$, and $b^k$ for $k\neq 
k_p,K$) we will use this numerical flux which is determined by the local 
numerical solutions. We also use this upwind form at a physical boundary (that
is, $a^1$ or $b^K$), but in this case a boundary condition supplies
the exterior solution.

Turning now to the endpoints $a^{k_p+1} = b^{k_p}$ corresponding 
to the particle location, we modify the standard upwind flux 
\eqref{eq:NumericalFlux} following the generalized 
discontinuous Galerkin method outlined in ~\cite{FanGDG}.
Consider a Dirac delta function located at the interface between 
elements $\mathsf{D}^{k_p}$ and $\mathsf{D}^{k_p + 1}$, and the
weak form of the resulting system \eqref{eq:firstordersys}.
The relevant new terms to consider have the form
\begin{align}
\label{DG_delta}
\int_{\mathsf{D}^{k_p}} d\xi
J_{1,2}\delta(\xi-\xi_p)\ell^{k_p}_j(\xi),\qquad
\int_{\mathsf{D}^{k_p + 1}} d\xi J_{1,2} \delta(\xi-\xi_p)
\ell^{k_p + 1}_j(\xi).
\end{align}
Upon evaluation, each if these terms appears similar in form 
to a boundary flux. The discontinuous Galerkin method method provides a 
self--consistent way to evaluate these integrals and then add the results 
to the numerical flux. We only require the usual selection property of the 
delta function when integrated over the {\em union} $\mathsf{D}^{k_p} \cup 
\mathsf{D}^{k_p + 1}$, and we are free to choose how the individual 
integrals over $\mathsf{D}^{k_p}$ and $\mathsf{D}^{k_p + 1}$ contribute 
to the total integral. In fact, the dynamics of
\eqref{eq:firstordersys} suggest a preferred distributional 
splitting. To see why, consider the scalar advection equation 
$(\partial_\lambda + v\partial_\xi) u = J(\xi,\lambda)\delta(\xi-\xi_p)$,
with $v>0$. Since this equation corresponds to rightward propagation,
the natural choice for the associated distributional splitting of
the delta function term is
\begin{align}
\int_{\mathsf{D}^{k_p}}d\xi J\delta(\xi-\xi_p)\ell^{k_p}_j(\xi)
= 0,\qquad
\int_{\mathsf{D}^{k_p + 1}}d\xi
J \delta(\xi-\xi_p)
\ell^{k_p + 1}_j(\xi) = J(\xi_p,t)\delta_{0,j}.
\end{align}
For this case, notice that the delta function only ``sees" a single
Lagrange polynomial, namely $\ell^{k_p + 1}_0(\xi)$ on the rightward
interval.

To enact an upwind splitting of the delta functions appearing the
system \eqref{eq:firstordersys}, we simply use the matrix $T^{-1}$
already defined in \eqref{eq:Udiagonalize} to isolate the two
propagating characteristic modes of the system. Consistent with
propagation of these modes, at the particle location we modify the 
fluxes given in Eqs.~(\ref{leftright_upwind_form}a,b),
\begin{subequations}\label{eq:NumericalFlux_PI1and2}
\begin{align}
\left(\begin{array}{c}(f_{\Pi}^{k_p + 1})^* \\(f_{\Phi}^{k_p + 1})^*
\end{array}\right)_{\text{left, modified}} & =
\left(\begin{array}{c}(f_{\Pi}^{k_p + 1})^* \\ (f_{\Phi}^{k_p + 1})^*
\end{array}\right)_\mathrm{left} +
T\left( \begin{array}{cc}1&0\\0&0 \end{array}\right)
T^{-1}\left(\begin{array}{c} J_1 \\ J_2 \end{array}\right) \\
\left(\begin{array}{c}(f_{\Pi}^{k_p})^*
\\(f_{\Phi}^{k_p})^* \end{array}\right)_{\text{right, modified}}
& =
\left(\begin{array}{c}(f_{\Pi}^{k_p})^* \\
(f_{\Phi}^{k_p})^* \end{array}\right)_\mathrm{right} +
T\left( \begin{array}{cc}
0 & 0 \\
0 & -1
\end{array}\right)
T^{-1}\left(\begin{array}{c} J_1 \\ J_2 \end{array}\right).
\end{align}
\end{subequations}
The correctness of this prescription can be see as follows. 
Integration of the system \eqref{eq:firstordersys} over the
union $\mathsf{D}^{k_p} \cup \mathsf{D}^{k_p + 1}$ followed 
by a subsequent integration by parts on each interval generates 
the following boundary terms at the particle location (and on 
the right--hand side of the equal sign):
\begin{equation}
\left.\left(\begin{array}{c}
f_{\Pi} \\
f_{\Phi}
\end{array}\right)\right|_{(\lambda,a^{k_p+1})}
-
\left.\left(\begin{array}{c}
f_{\Pi} \\
f_{\Phi}
\end{array}\right)\right|_{(\lambda,b^{k_p})}
+ 
\left(\begin{array}{c} 
J_1 \\ 
J_2
\end{array}\right).
\end{equation}
The two physical fluxes in this equation of course cancel each
other out, leaving only the vector $(J_1,J_2)^T$. Our modifications
(\ref{eq:NumericalFlux_PI1and2}a,b) of the numerical flux are
tailored to mimic this result. While the difference of the
left/right numerical fluxes at the particle location will not,
in general, cancel each other out (due to numerical error), notice 
that by subtracting (\ref{eq:NumericalFlux_PI1and2}b) from 
(\ref{eq:NumericalFlux_PI1and2}a) we generate precisely the vector 
$(J_1,J_2)^T$. This argument can be made more rigorous through an 
analysis based on integrating the two local numerical solutions on 
$\mathsf{D}^{k_p}$ and $\mathsf{D}^{k_p + 1}$ against the Lagrange 
polynomials $\ell^{k_p}_N(\xi)$ and $\ell^{k_p+1}_0(\xi)$. Finally, 
using the general expressions (\ref{eq:NumericalFlux}a,b), we may 
likewise succinctly express the modified numerical flux at the 
particle location as
\begin{equation}\label{eq:unifiedmodifiedflux}
\left(
\begin{array}{c}
(f_{\Pi}^{k})^* \\(f_{\Phi}^{k})^*
\end{array}\right)_{\text{modified}} 
= \left(\begin{array}{c}(f_{\Pi}^{k})^* \\ (f_{\Phi}^{k})^*
\end{array}\right) + \frac{1}{2}
T\left( \begin{array}{cc} 
1 - \mathbf{n}^- & 0
\\
0 & - 1 - \mathbf{n}^- \end{array}\right)
T^{-1}\left(\begin{array}{c} J_1 \\ J_2 \end{array}\right),
\end{equation}
where either $k = k_p$ or $k  = k_{p}+1$ in this equation.

\subsection{Initial data and boundary conditions}
The issues of initial data and boundary conditions are not
part of the dG method {\em per se}, but we must
nevertheless specify both to complete our numerical scheme. We
adopt trivial (zero) initial data, and avoid the issue of an
impulsively started problem by smoothly ``switching on" the source
terms, as discussed in Appendix \ref{sec:SourceDetails}.
At the boundaries we impose outgoing radiation boundary conditions.
Both potentials (\ref{eq:ZerilliV},\ref{eq:ReggeWheelerV}) behave
differently in the $\xi \rightarrow -\infty$ and $\xi \rightarrow
\infty$ limits, whence we treat the cases $\xi=a$ and $\xi=b$
differently. Since $1-2Mr^{-1} = 2Mr^{-1}\exp(-r/(2M))
\exp(x/(2M))$, in the $x \rightarrow -\infty$, $r \rightarrow 2M^{+}$
limit both potentials are exponentially small. Therefore, with
$a$ being sufficiently negative, $|V^\textrm{RW,Z}(r)|$ is zero to
machine precision when $r$ corresponds to $\xi \simeq a$, and
as an excellent approximation we may use the Sommerfeld
boundary condition
\begin{equation}
(\partial_t\Psi -\partial_x\Psi)(\lambda,a) = 0
\rightarrow \Pi(\lambda ,a)+\Phi(\lambda ,a)/x_\xi(\lambda,a)=0.
\end{equation}
In the $x,r \rightarrow \infty$ limit, both the Zerilli and
Regge--Wheeler potentials (\ref{eq:ZerilliV},\ref{eq:ReggeWheelerV})
behave like $V^\mathrm{RW,Z} = \ell(\ell+1)r^{-2} + O(r^{-3})$.
Therefore, were we to adopt a naive Sommerfeld condition at
$\xi = b$, the slow fall--off of the potential would
corrupt the benefits of our high--order accurate method. Instead,
we implement the radiation boundary condition described in \cite{LAU},
\begin{align}\label{eq:ROBC}
-\Pi(\lambda,b) + \Phi(\lambda,b)/x_\xi(\lambda,b) =
\frac{f(r_b)}{r_b}\int^\lambda_0
\Omega_\ell^\mathrm{RW,Z}(\lambda-\lambda',r_b)
\Psi(\lambda',b)d\lambda',
\end{align}
where $r_b = r(x(\lambda,b)) = r(b)$ and $\Omega_\ell^\mathrm{RW,Z}$
is a time--domain boundary kernel. As indicated, this kernel is
different for the Regge--Wheeler (here spin--2) and Zerilli cases,
although we suppress this dependence wherever possible.
\begin{table}
\subtable{
\scriptsize
\begin{tabular}{lcc}
\hline
\\
$k$
& Re$\beta{}^\mathrm{RW}_{2,k}(500)$
& Im$\beta{}^\mathrm{RW}_{2,k}(500)$ \\
\hline
 1   & $\hspace{5mm}$ $-$1.25849067540E$-$02 $\hspace{5mm}$ & 0 \\
 2   &     $-$8.23918644025E$-$03 &      0 \\
 3   &     $-$5.49064917188E$-$03 &      0 \\
 4   &     $-$3.62410271081E$-$03 &      0 \\
 5   &     $-$2.32805739548E$-$03 &      0 \\
 6   &     $-$1.42584745587E$-$03 &      0 \\
 7   &     $-$8.04688157035E$-$04 &      0 \\
 8   &     $-$3.83719341654E$-$04 &      0 \\
 9   &     $-$2.99532499571E$-$03 &      1.73407822255E$-$03\\
 & &
\\
\hline
\\
$k$ & Re$\gamma^\mathrm{RW}_{2,k}(500)$ & Im$\gamma^\mathrm{RW}_{2,k}(500)$
\\
\hline
 1 &       $-$8.36957985819E$-$09 &       0 \\
 2 &       $-$2.95922379193E$-$07 &       0 \\
 3 &       $-$2.97720676842E$-$06 &       0 \\
 4 &       $-$8.13540247121E$-$06 &       0 \\
 5 &       $-$1.40566197350E$-$06 &       0 \\
 6 &       $-$5.02202428400E$-$08 &       0 \\
 7 &       $-$1.01094068265E$-$09 &       0 \\
 8 &       $-$7.70486047714E$-$12 &       0 \\
 9 &       $-$2.99056309897E$-$03 &       1.73610608573E$-$03
\\
& &
\end{tabular}
}
\subtable{$\hspace{0.5cm}$}
\subtable{
\scriptsize
\begin{tabular}{lcc}
\hline
\\
$k$
& Re$\beta{}^\mathrm{Z}_{2,k}(500)$
& Im$\beta{}^\mathrm{Z}_{2,k}(500)$ \\
\hline
 1 & $\hspace{5mm}$ $-$1.25789030971E$-$02 $\hspace{5mm}$ & 0 \\
 2 &       $-$8.23529461921E$-$03 &       0 \\
 3 &       $-$5.48806353366E$-$03 &       0 \\
 4 &       $-$3.62239165593E$-$03 &       0 \\
 5 &       $-$2.32695433490E$-$03 &       0 \\
 6 &       $-$1.42517041551E$-$03 &       0 \\
 7 &       $-$8.04304980721E$-$04 &       0 \\
 8 &       $-$3.83535015275E$-$04 &       0 \\
 9 &       $-$2.99383340672E$-$03 &        1.73321233868E$-$03\\
 & &
\\
\hline
\\
$k$ & Re$\gamma^\mathrm{Z}_{2,k}(500)$ & Im$\gamma^\mathrm{Z}_{2,k}(500)$
\\
\hline
 1 &       $-$8.35513276685E$-$09 &       0 \\
 2 &       $-$2.95425498144E$-$07 &       0 \\
 3 &       $-$2.97239482588E$-$06 &       0 \\
 4 &       $-$8.12342297064E$-$06 &       0 \\
 5 &       $-$1.40379108037E$-$06 &       0 \\
 6 &       $-$5.01539234399E$-$08 &       0 \\
 7 &       $-$1.00959570760E$-$09 &       0 \\
 8 &       $-$7.69439666825E$-$12 &       0 \\
 9 &       $-$2.98758843820E$-$03 &        1.73437449497E$-$03
\\
& &
\end{tabular}
}
\caption{
\label{tab:kernels500}
Compressed kernels for $\ell =  2$, $r_b/(2M) =  500$,
$\varepsilon = 10^{-10}$. There are $d = 10$
poles and strengths, and complex conjugation of the ninth
entries gives the tenth entries. Zeros correspond to outputs
from the compression algorithm which are less than $10^{-30}$
in absolute value.
}
\end{table}

We approximate the time--domain boundary kernel 
$\Omega_\ell\simeq \Xi_\ell$ as a sum of exponentials
\begin{equation}
\Xi_\ell(t,r_b) = 
\sum_{k=1}^d \Xi_{\ell,k}(t,r_b),
\qquad
  \Xi_{\ell,k}(t,r_b) 
= \frac{\gamma_{\ell,k}(r_b/(2M))}{2M} 
\exp\left(\frac{t\beta_{\ell,k}(r_b/(2M))}{2M}\right).
\end{equation}
The parameters $\gamma_{\ell,k}(r_b/(2M))$ 
and $\beta_{\ell,k}(r_b/(2M))$ determine the
approximation $\Xi_{\ell}(t,r_b)$, and they depend on the 
Regge--Wheeler or Zerilli case, the orbital index $\ell$, 
and the dimensionless boundary radius $r_b/(2M)$. The
approximation $\Xi_\ell$ is designed so that its Laplace
transform agrees with the transform of $\Omega_{\ell}$
to relative supremum error $\varepsilon$ along the axis of 
imaginary Laplace frequency, and the the parameters
$\gamma_{\ell,k}$ and $\beta_{\ell,k}$
are the outputs from the Alpert--Greengard--Hagstrom
compression algorithm \cite{AGH,LAU}. Theoretically,
$\varepsilon$ is a long--time bound on the relative
convolution error in the time domain, and it measures 
the accuracy of the boundary condition. Table 
\ref{tab:kernels500} collects the $\ell=2$ kernels for 
$r_b = 1000M$ and $\varepsilon = 10^{-10}$. We evolve the 
constituent pieces of the approximate convolution via 
temporal integration of the ODE
\begin{equation}
\frac{d}{d\lambda}\int^\lambda_0
\Xi_{\ell,k}(\lambda-\lambda',r_b)
\Psi(\lambda',b)d\lambda'
= \frac{\beta_{\ell,k}}{2M}
\int^\lambda_0
\Xi_{\ell,k}(\lambda-\lambda',r_b)
\Psi(\lambda',b)d\lambda' + \Xi_{\ell,k}(0,r_b)
\Psi(\lambda,b),
\end{equation}
carrying out the integration along side, and coupled with, 
the numerical evolution of the system \eqref{eq:firstordersys}.
With this boundary condition, we are free to 
choose essentially any boundary 
$\xi=b$, so long as it lies to the right of the source. 
Our outer radiation boundary condition is especially useful when
studying eccentric orbits, for which one must average quantities
over many periods. 

%
%
\section{Numerical experiments and results}\label{sec:results}
\begin{figure}
\centering
\includegraphics[height=4in]{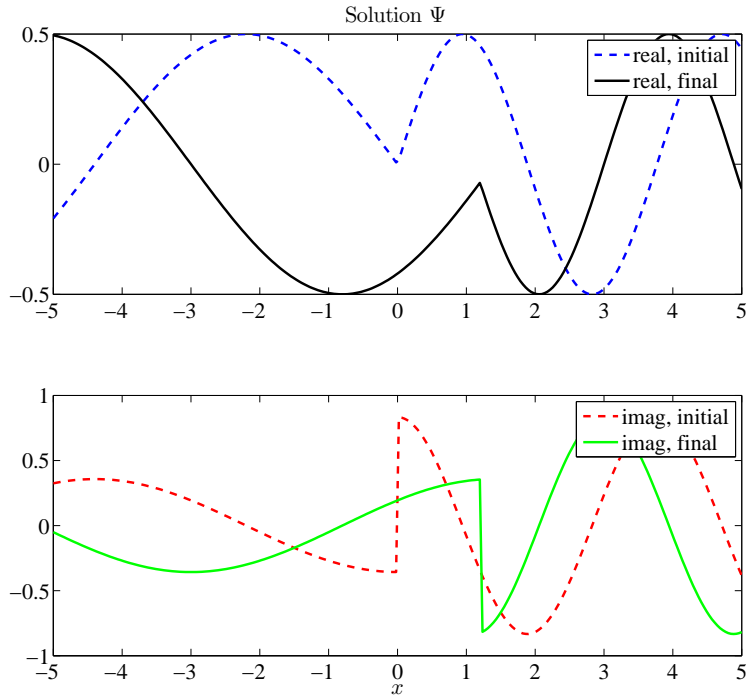}
\caption{{\sc $\Psi$--component of the solution.} The $\Pi$
and $\Phi$ components are qualitatively similar.}
\label{fig:MovingPsi}
\end{figure}

\subsection{Forced ordinary 1+1 wave equation}
We consider two scenarios involving an exact solution 
of the distributionally forced 1+1 wave equation with no 
potential. The first adopts exact initial data, and the 
second trivial data in parallel with the blackhole perturbation 
problem (a setting where trivial data is often chosen).

\subsubsection{Wave equation with exact initial data}
For a fixed velocity $v$ obeying $|v| < 1$, we consider
\begin{align}\label{eq:forced1p1}
-\partial_t^2\Psi + \partial_x^2\Psi 
=            \cos t\delta(x-vt) 
+ \mathrm{i} \cos t\delta'(x-vt).
\end{align}
Closed--form exact solutions 
for the real and imaginary parts of $\Psi$ are given in Appendix 
\ref{sec:WaveApp}, and we will check the convergence of our numerically 
generated solution against the complexification of these exact solutions. 
After expressing \eqref{eq:forced1p1} as a first order system and adopting 
our dG scheme, we obtain the same equations as in \eqref{eq:vectorsystem}, 
except now with a zero potential vector $\boldsymbol{V}$. Our domain is 
comprised of two subdomains: $\mathsf{D}^1$ to the left of the 
particle location $x_p(t) = vt$, and $\mathsf{D}^2$ to the right of 
$x_p(t)$. At $x_p(t)$, the interface between $\mathsf{D}^1$ and 
$\mathsf{D}^2$, we use Eq.~\eqref{eq:unifiedmodifiedflux} for
the numerical fluxes $(f_{\Pi}^k)^*$ and $(f_{\Phi}^k)^*$. At the 
physical boundary points we choose fluxes which enforce simple 
Sommerfeld boundary conditions,
\begin{equation}\label{eq:SommerForWave}
\Pi(\lambda,a)+\Phi(\lambda,a)/x_\xi(\lambda,a) = 0,\qquad 
\Pi(\lambda,b)-\Phi(\lambda,b)/x_\xi(\lambda,b) = 0.
\end{equation}
For this first experiment, we take initial data from the exact solution.
\begin{figure}
\centering
\includegraphics[height=4in]{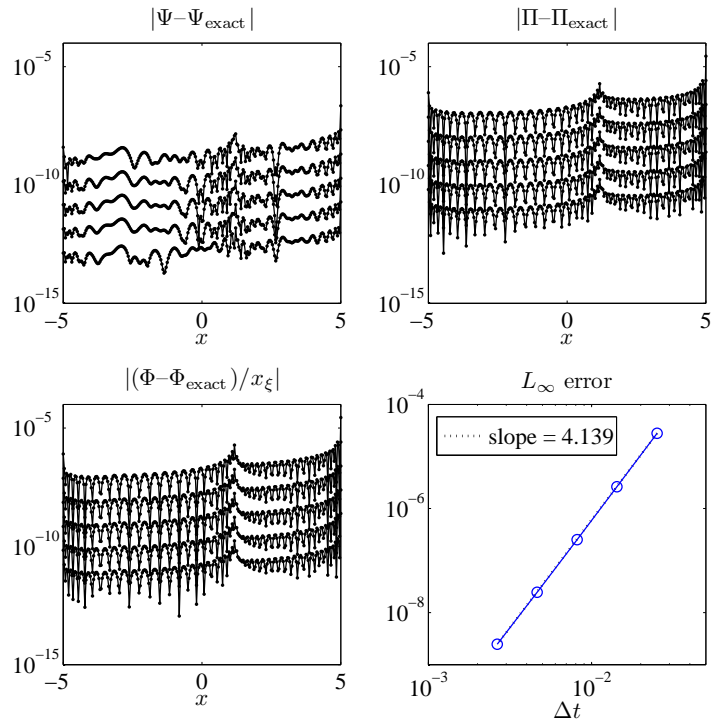}
\caption{{\sc Temporal convergence of the linearly moving
particle experiment.}
Errors have been computed relative to a uniformly spaced $x$--grid and
over all fields. The dotted line is a least--squares
fit of the data points (the round circles).}\label{fig:MovingError}
\end{figure}

\begin{figure}
\centering
\includegraphics[height=3.5in]{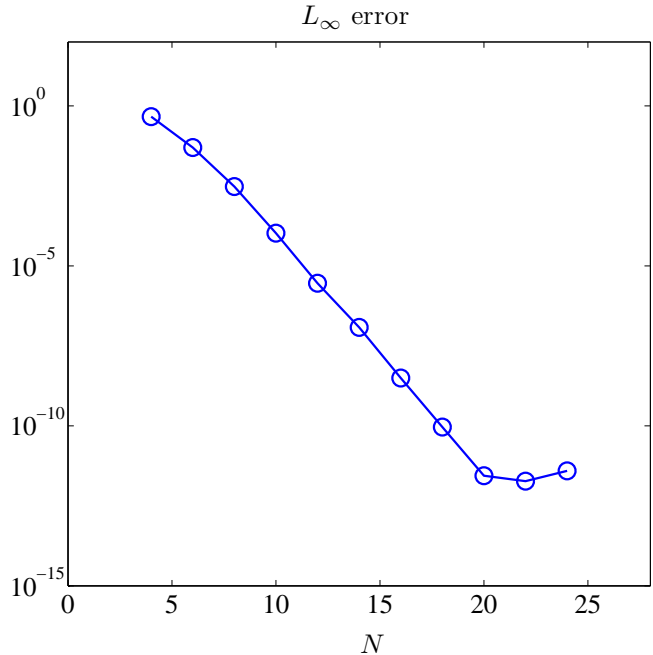}
\caption{{\sc Spectral convergence of the linearly moving particle
experiment.} Again, errors have been computed relative to a uniformly
spaced $x$--grid and over all fields.}\label{fig:ExpConvMovingParticle}
\end{figure}

Working with the global domain $[a,b] = [-5,5]$, we choose
$v = 0.4$ and the final time $t_F = 3.0$. For these choices the 
critical $\xi$ value \eqref{eq:criticalxi} always lies outside 
of the global domain, although clearly the example becomes 
pathological for a final time $t_F$ near $12.5$ (when the 
particle crosses the outer boundary). Fig.~\ref{fig:MovingPsi} 
shows the $\Psi$ component of the solution vector, and the 
$\Phi$ and $\Pi$ components also feature moving discontinuities. 
Fig.~\ref{fig:MovingError} documents the accuracy after several 
evolutions, each with $N = 26$ points, performed with 
decreasing temporal resolution in order to exhibit the 
fourth--order accuracy of the temporal Runge--Kutta integration. 
To compute errors, we have used the polynomial representations 
of the two local solutions, each computed with respect to the 
coordinates $(\lambda,\xi)$, to interpolate onto a uniformly 
spaced $x$--grid with 256 points where $L_\infty$ errors 
have been calculated. 
Fig.~\ref{fig:ExpConvMovingParticle} demonstrates 
the spectral convergence of our method for this problem. Here
$N$ is the number of points on each of the two subdomains, 
and for each $N$ we have chosen a $\Delta t$ to ensure 
stability.

\subsubsection{Wave equation with trivial initial data}
In this scenario we again consider \eqref{eq:forced1p1},
although now choosing $v = 0$ (a fixed particle analogous
to a circular orbit) and trivial initial data
$\Psi = \Pi = \Phi = 0$. As before, we will compute errors 
over all fields, after interpolation onto a reference grid, 
and against the exact solution. Before computing errors for the 
$\Psi$ variable when adopting trivial initial data, we have found 
it necessary to adjust the mean of our numerically generated 
$\Psi$ in order to ensure that it equals the mean of the exact 
$\Psi$. For the problem \eqref{eq:forced1p1} with Sommerfeld 
boundary conditions \eqref{eq:SommerForWave}, the $\Psi$ component 
of the solution vector is only determined up to an additive 
constant. Due to the presence of the potential, no such ambiguity 
is associated with solving \eqref{eq:genericwaveeq}.

\begin{figure}
\centering
\includegraphics[height=4in]{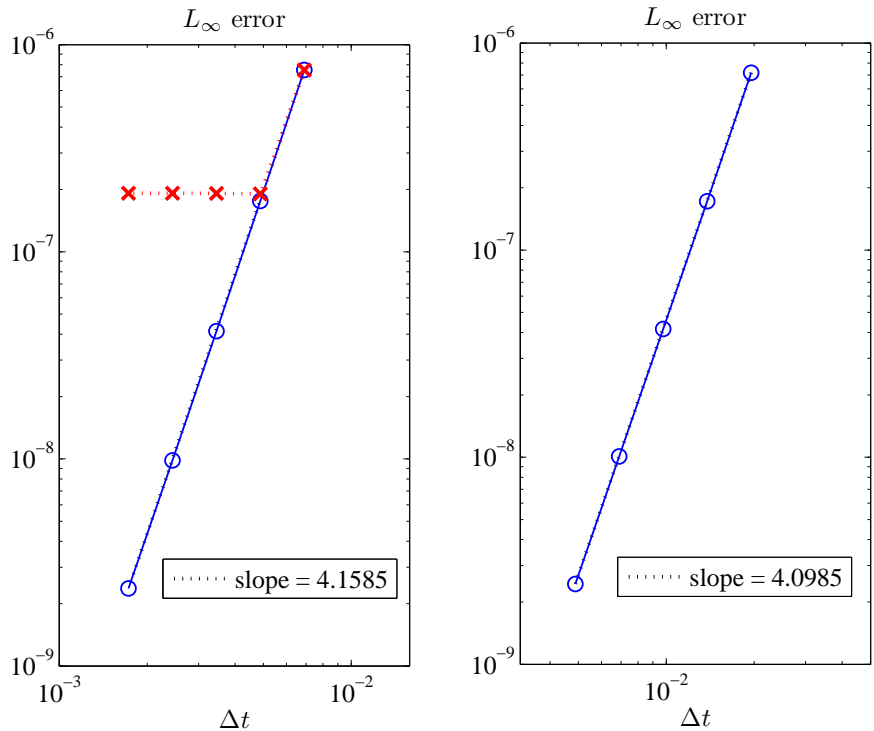}
\caption{{\sc Temporal convergence with trivial initial data.}
The left panel compares the two--domain experiment
run with and without the smoother, denoted by circles and crosses
respectively. The right panel corresponds to multiple
subdomains and no smoother. As described in the text, on each subdomain
we have fixed the mean of the numerical $\Psi$ to the exact value
before computing errors.}
\label{fig:ComparisonErrorSmoother}
\end{figure}

Our first test involves the minimal two domain set up. Since the 
problem is now impulsively started, we smooth the source functions 
as described in Eq.~\eqref{eq:smoothFandG}, choosing $t_0 = 0$, 
$\tau = 3$, and $\delta = 10$. For these choices, the source is 
smoothly ``switched on" (to machine precision) and is fully on by 
$t = 3$. Resolution of the transition requires relatively many 
points, and we have chosen $N = 61$ on each subdomain. For the 
final time $t_F = 10$, we demonstrate temporal convergence in
the left panel of Fig.~\ref{fig:ComparisonErrorSmoother}. 
We note that, as indicated in the figure, convergence is abruptly 
lost without the smoother. However, even without the smoother, 
by adopting multiple subdomains we also recover convergence to the 
exact solution (of course assuming $t_F > 5$, so that the initial 
incorrect profiles can fully propagate off the domain). Indeed,
the right panel of Fig.~\ref{fig:ComparisonErrorSmoother} 
documents the results for the same problem, but now without smoothing 
and 20 subdomains, each with $N = 7$ points. 
We explain this observation by noting that for $N = 1$ 
our dG method formally becomes a FV method. Therefore, many 
low--order elements corresponds to a more dissipative numerical 
flux, and the extra dissipation smooths the oscillations stemming 
from our impulsively started problem.

\subsection{Blackhole perturbations}

\subsubsection{Experiment: Zerilli equation with radiation 
boundary conditions}
This experiment involves the $\ell = 2$, $m = 2$ polar problem and a 
circular orbit with $p = 7.9456$, $M = 1$, and $m_p = 1$. We choose 
trivial initial data at $t_0 = 0$, with a smoother defined by $\tau 
= 10$ and $\delta = 10$. Integrating to final time $t_F = 90$, we 
first generate an accurate reference solution $\Psi_\mathrm{ref}$ on 
the domain $[-100,100]$, using 65+55 subdomains (65 to the left of
the particle and 55 to the right) with $N = 37$ nodal points on each. Here 
and below, we choose the time step $\Delta t$ to ensure stability.
At both endpoints $x = \pm 100$ we place Sommerfeld boundary 
conditions on $\Psi_\mathrm{ref}$, as physically no radiation 
reaches the endpoints by the final time.

The experiment is to generate a second numerical solution $\Psi$ 
on the shorter domain $[-50,b]$, where $b = 30 + 2\log(15-1) \simeq 
35.2$. We again evolve to final time $t_F = 90$, now with the 
convolution radiation boundary condition (\ref{eq:ROBC}) placed at 
the outer endpoint $x = b$. The relevant Zerilli kernel is defined 
in Table II of Ref.~\cite{LAU}. This kernel corresponds to $r_b/(2M) 
= 15$ and the tolerance $\varepsilon = 10^{-10}$. At the inner endpoint 
$x = -50$ we again adopt a Sommerfeld boundary condition. For
30+15 subdomains with 33 points on each, the corresponding $\Psi$ is 
then compared against the reference solution $\Psi_\mathrm{ref}$
in the $L_\infty$ norm. After interpolation onto a uniformly spaced 
grid with 853 points, we have found that 
$\|\Psi - \Psi_\mathrm{ref}\|_\infty \simeq 8.2314 \times 10^{-12}$.

\subsubsection{Results: circular orbits}

This subsection compares our numerical results for circular orbits to 
those obtained by other authors. For brevity we restrict ourselves to 
$\ell = 2$, but note that our method maintains its performance for 
higher $\ell$. We have considered higher $\ell$ values, and a fuller 
compilation of our results will appear elsewhere. For our simulations, 
we have chosen\footnote{By dividing Eq.~(\ref{eq:genericwaveeq}) by $m_p$ 
we can solve for the per--particle--mass perturbation $\Psi/m_p$ (from the 
coding standpoint, this is equivalent to setting $m_p = 1$). Physical 
waveforms and other quantities can then be recovered via multiplication by 
appropriate powers of $m_p$.}
$M = 1 = m_p$, with $\xi_\mathrm{max}=x_\mathrm{max}=1000+2\log(500-1) 
\simeq 1012$ and $\xi_\mathrm{min}=x_\mathrm{min}=-200$ as the outer and 
inner boundaries. We have used 45+200 subdomains, each with $N=21$ 
points, and a smoother \eqref{eq:smoothFandG} defined by $\tau = 1000$ 
and $\delta = 0.0002$. For these choices, we have integrated to $t_F = 
2500$ with time step $\Delta t = 0.005$. With these 
parameters we compute waveforms with a relative error of better 
than $10^{-8}$. Radiation boundary 
conditions \eqref{eq:ROBC} have been enforced through Table 
\ref{tab:kernels500}. Other parameters or non--uniformly placed 
subdomains may prove advantageous, but we have not explored all 
possibilities.

We first describe what we have measured. The luminosities of gravitational 
energy and angular momentum across an arbitrarily large spherical surface are 
determined from the master functions $\Psi^\mathrm{CPM}_{\ell m}(u+x,r)$
and $\Psi^\mathrm{Z}_{\ell m}(u+x,r)$. We view the retarded time
$u = t - x$ as fixed, but with $r,x$ arbitrarily large. Note that
$x \sim r$, as $r \rightarrow \infty$. In the $r \rightarrow \infty$ 
limit we have the energy and angular momentum luminosities across an 
infinite--radius spherical surface given by (see 
\cite{Martel_CovariantPert,SopuertaLaguna} and references therein)
\begin{subequations}\label{eq:EandLluminosity}
\begin{align}
\dot{E} = \sum_{\ell \geq 2} \sum_{m= -\ell}^\ell
\dot{E}_{\ell m},\qquad
\dot{E}_{\ell m} & =\frac{1}{64\pi}\frac{(\ell+2)!}{(\ell-2)!}
           \big(\big\vert\dot{\Psi}{}^\mathrm{CPM}_{\ell m}\big\vert^2 +
           \big\vert\dot{\Psi}{}_{\ell m}^\mathrm{ZM}\big\vert^2\big)\\
\dot{L} = \sum_{\ell \geq 2} \sum_{m= -\ell}^\ell
\dot{L}_{\ell m},\qquad
\dot{L}_{\ell m} & = \frac{\mathrm{i}m}{64\pi}
           \frac{(\ell +2)!}{(\ell -2)!}\big(
           \bar{\Psi}{}^\mathrm{CPM}_{\ell m}
           \dot{\Psi}{}^\mathrm{CPM}_{\ell m}
          +\bar{\Psi}{}^\mathrm{ZM}_{\ell m}
           \dot{\Psi}{}^\mathrm{ZM}_{\ell m}\big).
\end{align}
\end{subequations}
Here the overbar and dot denote complex conjugation and 
time differentiation respectively. The individual multipole contributions 
($\dot{E}_{\ell m}$ and  $\dot{L}_{\ell m}$) to the total energy and angular momentum 
luminousities decay exponentially with $\ell$ \cite{Martel_GravWave,Poisson1993,DRPP1971}. 
A few simplifications 
concerning $\dot{E}_{\ell m}$ and $\dot{L}_{\ell m}$ are 
worth noting. First, due to the fact that the particle 
moves in the equatorial plane, the following conditions hold: $\ell+m$ even 
$\implies \Psi^\mathrm{CPM} = 0$ and $\ell+m$ odd $\implies \Psi^\mathrm{ZM} = 0$. 
To establish these conditions, note, for example, that when $\ell+m$ is even the 
axial source terms $F^\mathrm{CPM}_{\ell m}$ and $G^\mathrm{CPM}_{\ell m}$ are 
identically zero. Second, from the behavior of the master functions under the 
mapping $m\rightarrow -m$, we have $\dot{E}_{\ell ,m}=\dot{E}_{\ell ,-m}$ and 
$\dot{L}_{\ell ,m}=\dot{L}_{\ell ,-m}$
\cite{Martel_GravWave}.

We stress that Eqs.~(\ref{eq:EandLluminosity}a,b) hold at infinity, 
and practically one must devise a way to extract the waveforms at infinity 
from the finite computation domain. This problem has been solved for 
waves on flat spacetime by Abrahams and Evans
\cite{AbrahamsEvans1988,AbrahamsEvans1990}. We will either simply 
``read off" waveforms at $r_\mathrm{max} = 1000$ or use Abrahams--Evans 
{\em flatspace} extraction. For $\ell = 2$ the procedure is as follows. 
We record a master scalar $\Psi$ at the outer boundary $x = b$ as a time 
series, and then integrate $\Psi(t,b) \simeq 
\ddot{f}(t-b) + 3\dot{f}(t-b)b^{-1} + 3f(t-b)
b^{-2}$ as if it were exact, thereby recovering the profile $f(t)$ and 
its derivatives. We perform a similar extraction on $\Pi$.
The Abrahams--Evans procedure is not exact for the perturbation equations 
we consider. Nevertheless, upon substitution of the approximate expansion 
$\ddot{f}(t-x) + 3\dot{f}(t-x)x^{-1} + 3f(t-x)x^{-2}$ into one of the
(homogeneous) master equations \eqref{eq:genericwaveeq}, we find a
residual which is $O\big(r^{-3}\log(\frac{1}{2}r/M)\big)$.
\begin{table}\scriptsize
\begin{tabular}{|c|c|c|c|c|c|c|}
\hline
\multicolumn{6}{|c|}{Energy
luminosity $(\dot{E}_{2m} + \dot{E}_{2,-m})/m_p^2$}    \\
\hline
\hline
$m$                            &
dG, read off                   &
dG, extract                    &
FE                             &
FR                             &
FD                             \\
\hline
1 & 
$8.17530620\times 10^{-7}$     &
$8.1633 \times 10^{-7}$        &
$8.1662 \times 10^{-7}$        &
$8.1633 \times 10^{-7}$        &
$8.1623 \times 10^{-7}$        \\
2 & 
$1.70685914 \times 10^{-4}$    &
$1.7062 \times 10^{-4}$        &
$1.7064 \times 10^{-4}$        &
$1.7063 \times 10^{-4}$        &
$1.7051 \times 10^{-4}$        \\
\hline
\multicolumn{6}{c}{}           \\
\multicolumn{6}{c}{}           \\
\hline
\multicolumn{6}{|c|}{Angular momentum
luminosity $(\dot{L}_{2m} + \dot{L}_{2,-m})/m_p^2$}    \\
\hline
\hline
$m$                            &
dG, read off                   &
dG, extract                    &
FE                             &
FR                             &
FD                             \\
\hline
1 & 
$1.83102416 \times 10^{-5}$    &
$1.8283 \times 10^{-5}$        &
$1.8289 \times 10^{-5}$        &
$1.8283 \times 10^{-5}$        &
$1.8270 \times 10^{-5}$        \\
2 & 
$3.82285415 \times 10^{-3}$    &
$3.8215 \times 10^{-3}$        &
$3.8219 \times 10^{-3}$        &
$3.8215 \times 10^{-3}$        &
$3.8164 \times 10^{-3}$        \\
\hline
\end{tabular}
\caption{{\sc $\ell = 2$ luminosities for a circular orbit with}
$(p,e)=(7.9456,0)$.
\label{table:circular_EL}
}
\end{table}
Table \ref{table:circular_EL} compares our dG, circular--orbit,
and $\ell = 2$ energy and angular momentum luminosities to results
obtained by other numerical methods described in the literature. 
Such a comparison is not straightforward as the 
finite--element (FE) results of Sopuerta and Laguna~\cite{SopuertaLaguna} 
involved reading off the master functions at $x=2000$, while 
the finite--difference (FD) results of Martel~\cite{Martel_GravWave} 
involved read--off at $x=1500$ (here we always assume $M = 1$).
The frequency--domain (FR) results of Poisson, as reported in
\cite{Martel_GravWave}, for the wave forms at infinity rely on 
the appropriate boundary value problems in the frequency domain, 
and of the three should afford the most direct comparisons.
\begin{table}\scriptsize
\begin{tabular}{|c|c|c|c|}
\hline
\multicolumn{4}{|c|}{Total $\ell = 2$ energy
luminosity $m_p^{-2}\sum_{m=-2}^2 \langle \dot{E}_{2m}\rangle$}   \\
\hline
\hline
Orbit parameters                                           &
dG, read off                                               &
dG, extract                                                &
FR                                                         \\
\hline
$e = 0.18891539$, $p = 7.50477840$                         &
$2.59367\times 10^{-4}$                                    &
$2.59296\times 10^{-4}$                                    &
$2.59296\times 10^{-4}$                                    \\
\hline
$e = 0.76412402$, $p = 8.75456059$                          &
$1.57146\times 10^{-4}$                                     &
$1.57120\times 10^{-4}$                                    &
$1.57131\times 10^{-4}$                                    \\
\hline
\multicolumn{4}{c}{}                                       \\
\multicolumn{4}{c}{}                                       \\
\hline
\multicolumn{4}{|c|}{Total $\ell = 2$ angular momentum luminosity
$m_p^{-2}\sum_{m=-2}^2 \langle \dot{L}_{2m}\rangle$}
\\
\hline
\hline
Orbit parameters                                           &
dG, read off                                               &
dG, extract                                                &
FR                                                         \\
\hline
$e = 0.18891539$, $p = 7.50477840$                         &
$4.91165\times 10^{-3}$                                    &
$4.91018\times 10^{-3}$                                    &
$4.91016\times 10^{-3}$                                    \\
\hline
$e = 0.76412402$, $p = 8.75456059$                         &
$2.09297\times 10^{-3}$                                    &
$2.09220\times 10^{-3}$                                    &
$2.09221\times 10^{-3}$                                    \\
\hline
\end{tabular}
\caption{{\sc Total $\ell = 2$ luminosities for eccentric orbits.}
\label{table:eccentric_luminosities}}
\end{table}
\begin{figure}
\centering
\includegraphics[height=3in]{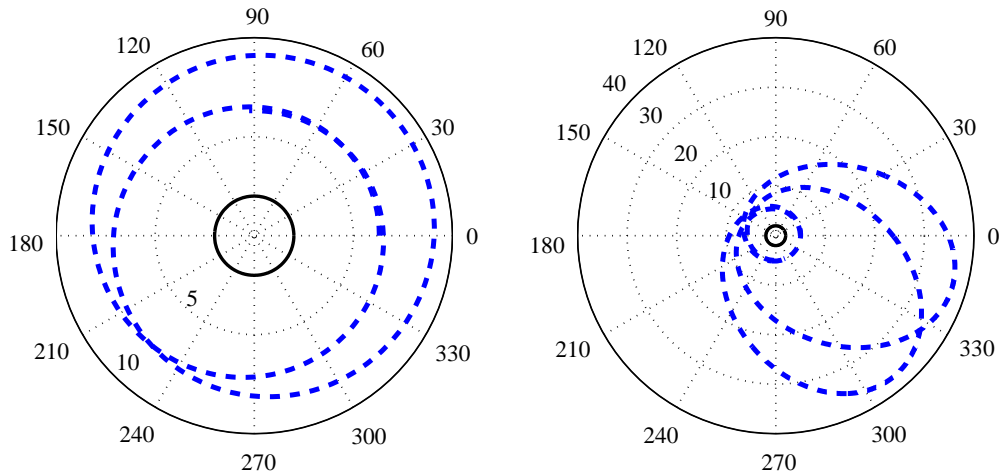}
\caption{{\sc Orbital paths.}  The left panel
shows one orbital period for $(e,p) =
(0.18891539,7.50477840)$. The right panel shows
two orbital periods for $(e,p) =
(0.76412402,8.75456059)$. In each case the dark inner circle
is the horizon. We have used the $(r,\phi)$
system to construct these polar plots.}
\label{fig:EccOrbit}
\end{figure}

\subsubsection{Results: eccentric orbits}
This subsection compares our numerical results for 
eccentric orbits to the frequency (FR) domain results of 
Tanaka {\em et al} \cite{SHIBATA} (rather than Poisson's 
frequency domain results). 
We again choose 45+200 subdomains, each with $N=21$ 
points, and $\Delta t \simeq 0.01$. Due to the incommensurate 
radial $T_r$ and azimuthal $T_\phi$ periods, 
we encounter the standard difficulty in obtaining measurements 
from eccentric--orbit simulations. Ideally, we would average 
measured luminosities over an infinite time, but will content 
ourselves with averaging over 4 radial cycles. Given a time 
series $A(t)$, we compute its corresponding average as
\begin{equation} \label{eqnAveragedQuant}
\langle A \rangle \equiv \frac{1}{T_2-T_1}\int_{T_1}^{T_2}dt A(t),
\qquad 
T_2 - T_1 = 4T_r.
\end{equation}

\begin{figure}
\centering
\includegraphics[height=3.5in, width=6.5in]{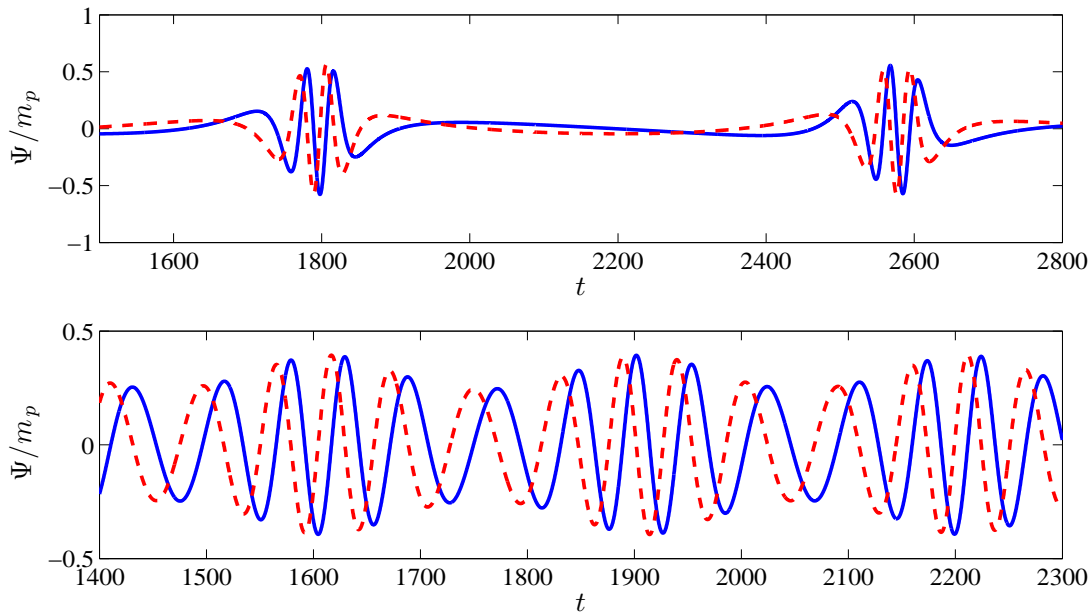}
\caption{{\sc Waveforms for $\ell = 2$, $m = 2$.}  
The top panel shows the $(e,p) = (0.76412402,8.75456059)$
extracted waveform, and the bottom panel 
the $(e,p) = (0.18891539,7.50477840)$ extracted waveform. 
Solid blue lines and dashed red lines respectively
correspond to real and imaginary parts.}
\label{fig:EccWaveforms}
\end{figure}

Table \ref{table:eccentric_luminosities} compares 
our total $\ell=2$ angular momentum and energy luminosities 
to the frequency (FR) domain results of Ref.~\cite{SHIBATA}.
In that reference the authors claim a relative numerical error 
of better than $10^{-4}$, which we have confirmed. We have 
retained enough significant digits in $(e,p)$ to match the 
parameters $(E_p,L_p)$ chosen in that reference. While we achieve 
relative errors of better than $10^{-4}$ for our {\em averaged} 
and {\em extracted} luminosities, we achieve single precision 
accuracy for our waveforms as a time series at $x = b$. 
Figure~\ref{fig:EccOrbit} exhibits the orbital paths for the 
two cases considered in this subsection, and 
Fig.~\ref{fig:EccWaveforms} shows the 
corresponding waveforms.

%
%
\section{Conclusion}
We have presented a high--order accurate discontinuous Galerkin
method for computing gravitational waveforms from extreme mass
ratio binaries. Time--domain approaches for computing such
waveforms have been hampered by the presence of distributional
source terms (which include both a moving Dirac delta function
and its derivative) in the governing master equations. By
writing a master equation as a first order system, we have 
treated the source term physically through an appropriate 
modification to the numerical flux function. Our method 
maintains spectral convergence without requiring additional 
procedures (e.g.~filtering), even pointwise in the immediate 
vicinity of the moving discontinuity. Through the use of 
convolution radiation boundary conditions, we have read--off 
waveforms at outer boundaries, thereby reducing computational 
cost without spoiling the high--order accuracy of our method. 
Accurate (read--off) waveforms, often with a relative error of 
better than $10^{-8}$, have been routinely observed in the 
course of our simulations.

This work has assumed that the particle trajectory is a 
timelike geodesic of the Schwarzschild geometry. However, the 
gravitational perturbations induced by the particle will, in 
turn, affect the particle trajectory. Several existing 
techniques capture this effect, thereby incorporating more
realistic inspiral (and possibly plunge) into the model.
These include gravitational self--force constructions
\cite{BMNOS2002,BarackSago,CanizaresSopuerta,Detweiler} 
(a representative, but far from exhaustive list),
as well as post--Newtonian calculations \cite{NDT2007} and 
adiabatic approximations \cite{ScottHughes_Adiabatic}.

We believe that the central ideas of our approach might apply
to many of these more sophisticated models. Computation of 
gravitational self--force requires that metric perturbations 
are regularized at the location of the particle. Mode--sum 
regularization has been carried out in the Lorenz gauge 
\cite{BarackSago}, in an approach where the metric perturbations 
are described by the full coupled system of 10 PDEs rather than 
the simpler master equation description. Although a dG approach 
might certainly be applied in this setting, our methods are 
directly applicable to approaches based on master equations. 
Retaining the master equation description, Detweiler 
has argued for regularizing gauge--invariant quantities 
\cite{Detweiler}. He has identified the appropriate quantities 
for quasi--circular orbits, and results based on these variables 
agree with corresponding Lorenz--gauge computations 
\cite{BarackDetComparison}. Our numerical method should prove ideal 
for self--force computations which require that the metric 
perturbations are well resolved near the particle. In particular, 
we hope to use our method in tandem with self--force corrections 
based on regularization of gauge--invariant quantities, at least 
for quasi--circular orbits. The post--Newtonian approach of 
Ref.~\cite{NDT2007} is also based on master equations, and we 
might also follow that reference in order to include 
nonconservative effects in our simulations.

In adapting our dG method to include, say, self--force effects we would 
surely encounter new difficulties. For example, due to dissipation 
associated with a self--force, the particle will inspiral, plunge, and 
merge with the blackhole. To handle all dynamical phases with 
our method, we would likely need to regrid at some point during 
the evolution. However, we believe this issue could be dealt with 
straightforwardly, given the robustness of the general method.

We conclude with remarks on the applicability of our dG method to 
perturbations of the Kerr metric. Here, we consider only the scenario of 
a particle following a timelike geodesic of the Kerr geometry, although
one might further consider gravitational self--force for this scenario 
as well. Now the relevant wave equation, the forced Teukolsky equation, 
is inherently 2+1 dimensional in the time--domain. In this case we would
need to ensure that the particle always lies on an edge between adjacent
subdomains (in this case triangles). Clearly, this is a geometrically
different problem, but Fan {\em et al}~\cite{FanGDG} have also
considered 2+1 problems, and one might pursue the Kerr problem along
similar lines.

\section*{Acknowledgments}
We would like to thank Jae--Hun Jung, Carlos Sopuerta, and 
Richard Price for correspondence and useful discussions, and 
gratefully acknowledge support through grants DMS 0554377 
and DARPA/AFOSR FA9550-05-1-0108 to Brown University.
\appendix

%
%
\section{Jump conditions}\label{AppJumps}
The derivation 
of the jump conditions \eqref{eq:Alljumps}
goes as follows. Using the 
selection properties of $\delta'(u)$ as a distribution, 
we first rewrite (\ref{eq:genericwaveeq}) as
\begin{align}
-\partial^2_t\Psi + \partial^2_x\Psi & -V(r)\Psi 
= f_p(t) F(t,r_p(t)) \delta'(r-r_p(t))
\nonumber \\
& + \big[f_p(t)G(t,r_p(t))
-g_p(t)F(t,r_p(t))
-f_p(t)F_r(t,r_p(t))\big]
\delta(r-r_p(t)),
\label{eq:distforceRHS}
\end{align}
using the shorthand notations from \eqref{eq:pshorthands}. 
Next, with $L$ for ``left'' and $R$ for ``right'', we let
\begin{equation}
\Psi(t,r) = \Psileft(t,r)\theta(r_p(t)-r)
+\Psiright(t,r)\theta(r-r_p(t)),
\label{eq:Psisplit}
\end{equation}
where the step function $\theta(u)$ obeys $\theta(u) = 0$ 
for $u < 0$, and $\theta(u) = 1$ for $u > 0$. We view the 
functions $\Psi^{L,R}$ as everywhere satisfying the 
homogeneous PDE
\begin{equation}
-\partial^2_t\Psi^{L,R} 
+\partial^2_x\Psi^{L,R} 
- V(r) \Psi^{L,R} = 0,
\label{eq:homog}
\end{equation}
even across the particle location $r_p(t)$.

To complete our derivation of (\ref{eq:Alljumps}), we 
calculate the distributional 
derivatives of $\Psi$ as given in (\ref{eq:Psisplit}), 
insert them into (\ref{eq:distforceRHS}), and then compare
terms. Using $\partial/\partial x = f\partial/\partial r$, 
the identity $\theta'(u) = \delta(u)$, and
the argument symmetry of $\delta(u)$,
we first compute
\begin{equation}
\partial_x\Psi = 
 \Psileft_x\theta(r_p(t)-r)
+\Psiright_x\theta(r-r_p(t))
- f\Psileft\delta (r-r_p(t))
+ f\Psiright\delta (r-r_p(t)),
\end{equation}
where on the right--hand side we have switched to subscript 
notation for partial derivatives. The second $x$--derivative 
of (\ref{eq:Psisplit}) is then
\begin{align}
& \partial^2_x\Psi = \Psileft_{xx}\theta(r_p(t)-r)
+\Psiright_{xx}\theta(r-r_p(t)) 
- 2f^2\Psileft_r\delta (r-r_p(t))
+ 2f^2\Psiright_r \delta (r-r_p(t))\nonumber \\
& \hspace{1.2mm} - f f'\Psileft\delta (r-r_p(t))
+ f f'\Psiright \delta (r-r_p(t)) 
- f^2\Psileft\delta' (r-r_p(t))
+ f^2\Psiright\delta' (r-r_p(t)).
\end{align}
Expressed compactly, the last formula is
\begin{align}
\partial^2_x\Psi & = \Psileft_{xx}\theta(r_p(t)-r)
+\Psiright_{xx}\theta(r-r_p(t))
+ f_p^2(t)\ljump\Psi_r\rjump \delta (r-r_p(t))
\nonumber \\
& - f_p(t) g_p(t) \ljump\Psi\rjump\delta (r-r_p(t))
+ f_p^2(t)\ljump\Psi\rjump\delta' (r-r_p(t)),
\label{eq:distPsixx}
\end{align}
where the definition (\ref{eq:jumpdefinition}) is here 
$\ljump\Psi\rjump(t) \equiv \Psiright(t,r_p(t)) - \Psileft(t,r_p(t))$.
To reach Eq.~(\ref{eq:distPsixx}) from the previous line,
we have used the selection properties of $\delta'(u)$. 
Next, we similarly compute 
\begin{align}
\partial^2_t\Psi & = \Psileft_{tt}\theta(r_p(t)-r)
+\Psiright_{tt}\theta(r-r_p(t)) 
+ 2\dot{r}_p\Psileft_t\delta (r-r_p(t))
- 2 \dot{r}_p\Psiright_t \delta (r-r_p(t))\nonumber \\
& + \ddot{r}_p\Psileft\delta (r-r_p(t))
- \ddot{r}_p\Psiright\delta (r-r_p(t)) 
- \dot{r}^2_p\Psileft\delta' (r-r_p(t))
+ \dot{r}^2_p\Psiright\delta' (r-r_p(t)).
\end{align}
The last formula may be written in the succinct form
\begin{align}
\partial^2_t\Psi & = \Psileft_{tt}\theta(r_p(t)-r)
+\Psiright_{tt}\theta(r-r_p(t)) 
- 2\dot{r}_p(t)\ljump\Psi_t\rjump\delta (r-r_p(t))
\nonumber \\
&
- \ddot{r}_p(t)\ljump\Psi\rjump\delta (r-r_p(t))
- \dot{r}^2_p(t)\ljump\Psi_r\rjump\delta(r-r_p(t))
+ \dot{r}^2_p(t)\ljump\Psi\rjump\delta' (r-r_p(t)),
\end{align}
again by using the properties of $\delta'(u)$.
Finally, with $\partial_t \ljump\Psi\rjump =
\ljump\Psi_t\rjump + \dot{r}_p(t) \ljump\Psi_r\rjump$, 
we rewrite the last expression as
\begin{align}
\partial^2_t\Psi & = \Psileft_{tt}\theta(r_p(t)-r)
+\Psiright_{tt}\theta(r-r_p(t))
- 2\dot{r}_p(t)\left(\partial_t\ljump\Psi\rjump\right)
\delta (r-r_p(t))
\nonumber \\
&
- \ddot{r}_p(t)\ljump\Psi\rjump\delta (r-r_p(t))
+ \dot{r}^2_p(t)\ljump\Psi_r\rjump\delta(r-r_p(t))
+ \dot{r}^2_p(t)\ljump\Psi\rjump\delta' (r-r_p(t)).
\label{eq:distPsitt}
\end{align}
Substitution of (\ref{eq:distPsixx}) and (\ref{eq:distPsitt}) into
(\ref{eq:distforceRHS}), along with the fact that $\Psiright$ and 
$\Psileft$ solve the homogeneous PDE (\ref{eq:homog}), 
then yields Eqs.~(\ref{eq:jumpPsi}) and (\ref{eq:jumpPsitPsiPsir}). 

\section{Exact solutions to the forced 1+1 wave equation}
\label{sec:WaveApp}
This appendix presents exact solutions to the distributionally 
forced 1+1 wave equation. Precisely, we consider the 
equation
\begin{equation}\label{eq:appendixWaveEq}
-\partial_t^2\Psi +\partial_x^2\Psi = G(t)\delta(x-vt) 
+ F(t)\delta'(x-vt),
\end{equation}
where either $F(t) = 0, G(t) = \cos t$ or
$F(t) = \cos t, G(t) = 0$. In analyzing both cases, we make 
use of the following distributional identities: 
\begin{equation}
\partial_u |u| = \sgn u,\qquad
\partial_u \sgn u = 2\delta(u),\qquad
(\sgn u)^2=1,
\label{eq:sgnidentities}
\end{equation}
with $\sgn u \equiv u/|u|$ the sign function. Throughout, the 
particle location $x_p(t) = vt$ has linear time dependence, 
with corresponding speed $|v| < 1$.

\subsection{Solution for $F(t) = 0$, $G(t) = \cos t$}
Following analysis similar to that presented in Section 
\ref{subsec:masterJumps} [or by substituting the 
correspondences $x_p(t) = r_p(t)$, $x = r$, $f(r) = 1$, 
$f'(r) = 0$, and $\ddot{r}(t) = 0$ into the general
jumps \eqref{eq:Alljumps}], we find the jump relations
\begin{equation}\label{eq:appendixJumpsSolnZeroF}
\ljump\Psi\rjump_{x=vt}=0,\qquad
\ljump\partial_x\Psi\rjump_{x=vt}=\gamma^2\cos t,\qquad
\ljump\partial_t \Psi\rjump_{x=vt}= -v\gamma^2\cos t,
\end{equation}
where $\gamma = (1-v^2)^{-1/2}$ is the usual relativistic
factor. The particular solution 
\begin{align}\label{eq:appendixSolnZeroF}
\Psi(t,x) = -{\textstyle \frac{1}{2}}\sin\vartheta,\qquad 
\vartheta = \gamma^2(t-xv-|x-vt|)
\end{align}
to Eq.~\eqref{eq:appendixWaveEq} possess the jumps listed 
in \eqref{eq:appendixJumpsSolnZeroF}.
Using the identities (\ref{eq:sgnidentities}),
let us verify that \eqref{eq:appendixSolnZeroF} indeed solves 
\eqref{eq:appendixWaveEq} for $F(t) = 0$ and $G(t) =\cos t$.
Straightforward computation of the first and second order
$t$--derivatives yields
\begin{align}
\partial_t \Psi  & = 
                 -{\textstyle \frac{1}{2}}
                 \gamma^2[1+v\sgn(x-vt)] 
                 \cos\vartheta\\
\partial_t^2\Psi & =  
                  {\textstyle \frac{1}{2}}
                  \gamma^4[1+v\sgn(x-vt)]^2
                  \sin\vartheta
                  +v^2\gamma^2\delta(x-vt)\cos\vartheta.
\end{align}
while for the $x$--derivatives we similarly find
\begin{align}
\partial_x\Psi   & =  
                 {\textstyle \frac{1}{2}}
                 \gamma^2[v+\sgn(x-vt)]
                 \cos\vartheta\\
\partial_x^2\Psi & =  
                 {\textstyle \frac{1}{2}}
                 \gamma^4 [v+\sgn(x-vt)]^2
                 \sin\vartheta
                 +\gamma^2\delta(x-vt)\cos\vartheta.
\end{align}
Forming $-\partial_t^2\Psi + \partial_x^2\Psi = \delta(x-vt) \cos\vartheta$, we then 
appeal to the selection property of the delta function in order to 
reach the desired result, $-\partial_t^2\Psi+\partial_x^2\Psi = 
\delta(x-vt) \cos t$.

\subsection{Solution for $F(t) = \cos t$, $G(t) = 0$}
The jump relations for this case are
\begin{equation}\label{eq:appendixJumpsSolnZeroG}
\ljump\Psi\rjump_{x=vt}=\gamma^2 \cos t,\quad
\ljump\partial_x\Psi\rjump_{x=vt}=2v\gamma^4\sin t,\quad
\ljump\partial_t \Psi\rjump_{x=vt}= -(1+v^2)\gamma^4\sin t.
\end{equation}
Now a particular solution to \eqref{eq:appendixWaveEq} is
\begin{align}\label{eq:appendixSolnZeroG}
\Psi(t,x) = {\textstyle \frac{1}{2}} \gamma^2
            [v+\sgn(x-vt)]\cos\vartheta,\qquad 
\vartheta = \gamma^2(t-xv-|x-vt|).
\end{align}
To verify that \eqref{eq:appendixSolnZeroG} indeed solves 
\eqref{eq:appendixWaveEq} for $F(t) = \cos t$ and $G(t) = 0$,
we first calculate
\begin{align}\label{eq:Psittcase2}
\partial_t \Psi & = -{\textstyle \frac{1}{2}}\gamma^4
                    [2v + (1+v^2)\sgn(x-vt)] 
                    \sin\vartheta 
                    -v\gamma^2
                    \delta(x-vt) \cos\vartheta \\
\partial_t^2\Psi & =v^2\gamma^2 \delta'(x-vt)\cos\vartheta
                   +\gamma^4 [2v+v^3+v^2\sgn(x-vt)]
                   \delta(x-vt)\sin\vartheta\nonumber \\
                 & -{\textstyle \frac{1}{2}}\gamma^6
                    [3v+v^3+(3v^2+1)\sgn(x-vt)]
                    \cos\vartheta,
\end{align}
and then likewise compute
\begin{align}\label{eq:Psixxcase2}
\partial_x\Psi    & =  \gamma^2\delta(x-vt)\cos\vartheta
                       +{\textstyle \frac{1}{2}}\gamma^4
                       [1+v^2+2v\sgn(x-vt)]
                       \sin\vartheta \\
\partial_x^2\Psi & =  \gamma^2 \delta'(x-vt)\cos\vartheta
                      +\gamma^4[3v+\sgn(x-vt)]
                      \delta(x-vt)\sin\vartheta
                      \nonumber \\
                      &-{\textstyle \frac{1}{2}}\gamma^6
                      [3v+v^3+(3v^2+1)
                      \sgn(x-vt)]\cos\vartheta.
\end{align}
Combination of Eqs.~(\ref{eq:Psittcase2}) and (\ref{eq:Psixxcase2}) 
yields
\begin{equation}
\label{eq:eqGzeroTemp}
-\partial_t^2\Psi+\partial_x^2\Psi= \delta'(x-vt)\cos\vartheta
                                   +\gamma^2[v+\sgn(x-vt)]
                                    \delta(x-vt)\sin\vartheta.
\end{equation}
By the selection properties of $\delta'(u)$, we have
\begin{equation}
\delta'(x-vt)
\cos\vartheta
= \delta'(x-vt) \cos t 
-\gamma^2[v+\sgn(x-vt)]
\delta(x-vt)\sin t.
\end{equation}
Substituting this result into \eqref{eq:eqGzeroTemp}, using
the selection property of $\delta(u)$, and realizing that
$\delta(u)\sgn u = 0$ by symmetry, we arrive at the 
desired result, $-\partial_t^2\Psi+\partial_x^2\Psi=
\delta'(x-vt)\cos t$.

\section{Source terms}\label{sec:SourceDetails}
This appendix lists the specific functions $F_{\ell m}(t,r)$ and
$G_{\ell m}(t,r)$ for our polar and axial cases. However, one
practical difference between the functions listed below, and what 
we have used in our numerical simulations is the following. In 
accordance with our unphysical choice of vanishing initial data, we 
``switch on" the source terms smoothly via the following prescription:
\begin{equation}\label{eq:smoothFandG}
F_{\ell m}(t,r) \rightarrow 
\left\{\begin{array}{rcl}
{\textstyle \frac{1}{2}}
[\mathrm{erf}(\sqrt{\delta}(t - t_0 - \tau/2)+1]
F_{\ell m}(t,r)
& & \text{for } t_0\leq t \leq t_0+\tau\\
F_{\ell m}(t,r) & & \text{for } t > t_0+\tau,
\end{array}\right.
\end{equation}
and the same for $G_{\ell m}(t,r)$. Typically, the initial time $t_0
= 0$, and the timescale $\tau$ is much shorter than the final time 
of the run. Choosing suitable $\tau$ and $\delta$, we ensure that 
our start-up is smooth to machine precision, and thereby avoid the 
troublesome nature of an impulsively started problem. Note that this 
prescription does initially affect the form of $\partial_t F_{\ell m}(t,r)$.

To both express and derive concrete expressions for the source terms,
we rely on standard results for particle motion in the Schwarzschild 
geometry \cite{MTW,Chan_MathBlack}. As an alternative set to $(e,p)$ 
discussed in Section \ref{subsec:particlemotion}, we may instead work 
with $(E_p,L_p)$, the physical particle energy and angular momentum 
(both per unit mass). These constants of the motion are related to our 
original set by \cite{CUT,Martel_GravWave}
\begin{align}
L_p^2 = \frac{p^2M^2}{p-3-e^2}, \qquad 
E_p^2 = \frac{(p-2)^2 - 4e^2}{p(p-3-e^2)}.
\end{align}
Let $z^\mu(\tau) = (t(\tau),r(\tau),\theta(\tau),\phi(\tau))$ be 
the parameterization of the particle's four--trajectory in terms of 
proper time $\tau$. As before, set $r_p(t) = r(\tau(t))$ for the radial 
coordinate of the particle expressed in terms of coordinate time, with 
similar expressions for $\theta_p(t)$ and $\phi_p(t)$. We assume that 
the coordinate system has been selected to ensure equatorial motion,
$\theta_p(t) = \pi/2$. Then the four--velocity $u^\mu = dz^\mu/d\tau$
has components
\begin{equation}\label{eq:fourvelocity}
u^t = E_p/f(r),\quad (u^r)^2 = E_p^2 -f(r)(1 +L_p^2/r^2), \quad
u^\theta = 0,\quad u^\phi = L_p/r^2,
\end{equation}
where these expressions follow from standard conservation arguments
\cite{Chan_MathBlack,MTW}. 

Throughout this appendix, we use $m_p$ for the particle mass in order
to avoid confusion with the azimuthal spherical harmonic index $m$.

\subsection{Zerilli--Moncrief (polar) source term}
To define the Zerilli--Moncrief source term, we first introduce the 
polar spherical harmonics
\begin{align}
Y^{\ell m},\quad
Y^{\ell m}_a = Y^{\ell m}_{:a},\quad
Y^{\ell m}_{ab}=Y^{\ell m}\gamma_{ab},\quad
Z^{\ell m}_{ab}= Y^{\ell m}_{:ab}
+\frac{\ell(\ell+1)}{2}Y^{\ell m}\gamma_{ab},
\end{align}
where $Y^{\ell m}(\theta,\phi)$ are the ordinary scalar harmonics 
\cite{ThorneMultipole}, $\gamma_{ab}$ is the metric of the 
unit--radius round sphere, and a colon indicates covariant
differentiation compatible with $\gamma_{ab}$. The Zerilli--Moncrief 
source term is specified by\footnote{Factors of $f(r)$ are included 
here in order to have direct comparison with the same coefficients 
listed in Refs.~\cite{SopuertaLaguna,Martel_GravWave}.}
\begin{subequations} \label{eq:polarFandG}
\begin{align}
f(r)F^\mathrm{ZM}_{\ell m}(t,r) & = 
      e_{\ell}(r)\bar{Y}^{\ell m}(t)\\
f(r)G^\mathrm{ZM}_{\ell m}(t,r)  & = 
      a_{\ell}(r)\bar{Y}^{\ell m}(t)
     +b_{\ell}(r)\bar{Y}^{\ell m}_\phi(t)
     +c_{\ell}(r)\bar{Y}^{\ell m}_{\phi\phi}(t)
     +d_{\ell}(r)\bar{Z}^{\ell m}_{\phi\phi}(t).
\end{align}
\end{subequations}
Here, for example, $\bar{Y}^{\ell m}(t) \equiv 
\bar{Y}^{\ell m}(\pi/2,\phi_p(t))$. Moreover, the coefficients in
\eqref{eq:polarFandG} are given by \cite{SopuertaLaguna,Martel_GravWave}
\begin{subequations}\label{eq:ZMcoefficients}
\begin{align}
a_{\ell}(r)  & =  \frac{8\pi m_p}{(1+n_\ell)} 
                  \frac{f^2(r)}{r\Lambda_\ell^2(r)}
                  \left\{\frac{6M E_p}{r}
                 -\frac{\Lambda_\ell(r)}{E_p}\left[1+
                  n_\ell-\frac{3M}{r}
                 +\frac{L_p^2}{r^2}
                  \left(n_\ell+3-\frac{7M}{r}\right)
                  \right]\right\}\\
b_{\ell}(r)  & =  \frac{16\pi m_p}{(1+n_\ell)} 
                  \frac{f^2(r)}{r^2\Lambda_\ell(r)}
                  \frac{L_p}{E_p}u^r\\
c_{\ell}(r)  & =  \frac{8\pi m_p}{(1+n_\ell)} 
                  \frac{f^3(r)}{r^3\Lambda_\ell(r)}
                  \frac{L_p^2}{E_p}\\
d_{\ell}(r)  & = -32\pi m_p
                 \frac{(\ell-2)!}{(\ell+2)!}
                 \frac{f^2(r)}{r^3}
                 \frac{L_p^2}{E_p}\\
e_{\ell}(r)  & = \frac{8\pi m_p}{(1+n_\ell)}
                 \frac{f^3(r)}{\Lambda_\ell(r)}
                 \frac{1}{E_p}
                 \left(1+\frac{L_p^2}{r^2}\right),
\end{align}
\end{subequations}
where $n_\ell = (\ell+2)(\ell-1)/2 = \Lambda_\ell(r) - 3M/r$, and 
$u^r$ is determined by \eqref{eq:fourvelocity} and the sign of 
$\dot{r}_p(t)$. Due to the $u^r$ factor, we may not,
strictly speaking, interpret $b_{\ell}(r)$ as solely a function
of $r$, but $f(r)u^r/E_p$ could also be reinterpreted as 
$\dot{r}_p(t)$ and paired with $\bar{Y}^{\ell m}_\phi(t)$.

\subsection{Cunningham--Price--Moncrief (axial) source term}
With $\epsilon_{ab}$ the unit--sphere Levi--Civita 
tensor such that $\epsilon_{\theta\phi} = -\sin\theta$, the 
axial spherical harmonics are
\begin{equation}
S^{\ell m}_a = \gamma^{bc}\epsilon_{ab}Y^{\ell m}_{:c},\qquad
S^{\ell m}_{ab} = S^{\ell m}_{(a:b)},
\end{equation}
and we express the Cunningham--Price--Moncrief source term as
\begin{subequations}\label{eq:CPMsource}
\begin{align}
f(r)F^\mathrm{CPM}_{\ell m}(t,r) & =C_{\ell}(r)\bar{S}^{\ell m}_\phi(t) \\
f(r)G^\mathrm{CPM}_{\ell m}(t,r) & =A_{\ell}(r)\bar{S}^{\ell m}_\phi(t) +
B_{\ell}(r)\bar{S}^{\ell m}_{\phi \phi}(t).
\end{align}
\end{subequations}
As before, $(t)$ indicates evaluation on $(\theta,\phi) 
= (\pi/2,\phi_p(t))$,
and the coefficients in the above expressions are as follows:
\begin{subequations}\label{eq:CPMcoefficients}
\begin{align}
A_{\ell}(r) & = 32\pi m_p \frac{(\ell-2)!}{(\ell+2)!} 
              \frac{f^2(r)}{r^2}\frac{L_p}{E_p^2}
              \left[
              f(r) - 2 E_p^2 -
               \left(1-\frac{5M}{r}\right)
              \left(1+\frac{L_p^2}{r^2}\right)
              \right]\\
B_{\ell}(r) & = 32\pi m_p \frac{(\ell-2)!}{(\ell+2)!} \frac{f^2(r)}{r^3}
\frac{L_p^2}{E_p^2} u^r\\
C_{\ell}(r) & = 32\pi m_p \frac{(\ell-2)!}{(\ell+2)!} \frac{f^3(r)}{r}
\frac{L_p}{E_p^2} \left(1+\frac{L_p^2}{r^2}\right).
\end{align}
\end{subequations}
As before, we may not truly interpret $B_{\ell}(r)$ as a function of $r$,
but nevertheless keep this convenient notation. We note that our 
$A_\ell(r)$ does not agree with the corresponding factor
$u_\ell(r)$ quoted in Ref.~\cite{SopuertaLaguna}; however, we find 
that $u_\ell(r) = A_\ell(r) - C_\ell'(r)$. Due to this discrepancy, 
we present our derivation of \eqref{eq:CPMcoefficients}.

\subsection{Derivation of the axial source term}\label{sec:CPM_der}
Our goal is to establish formulas 
(\ref{eq:CPMsource},\ref{eq:CPMcoefficients}) for the
Cunningham--Price--Moncrief source term,
\begin{equation}\label{eq:PoissonCPM}
S^\mathrm{CPM} = G^\mathrm{CPM}\delta(r-r_p(t))
                +F^\mathrm{CPM}\delta'(r-r_p(t)).
\end{equation}
Here and in what follows, we suppress $(\ell,m)$ indices wherever 
possible. Our starting point is Martel and Poisson's expression 
for $S^\mathrm{CPM}$ in $(t,r)$ coordinates,
\begin{equation}\label{eq:S_CPM}
S^\mathrm{CPM} = \frac{2r}{(\ell-1)(\ell+2)}
                 \left(f^{-1}\partial_t P^r+f\partial_r
                 P^t+\frac{2M}{r^2}P^t\right).
\end{equation}
This result appears in Appendix C of their expanded version of 
\cite{Martel_CovariantPert}, where $S_\mathrm{odd}$ in that 
reference is our $S^\mathrm{CPM}$. The vector $P^A = (P^t,P^r)$ 
is given by Eq.~(5.10) of \cite{Martel_CovariantPert},
\begin{align}\label{eq:PA}
P^A= \frac{16\pi r^2}{\ell(\ell+1)}\int_{S^2} 
     d\Omega T^{Ab}\bar{S}_b^{\ell m},
\end{align}
but here with our index conventions and harmonics. Note that 
$S_a^{\ell m}$ corresponds to $X_A^{\ell m}$ of 
\cite{Martel_CovariantPert}.

The stress--energy tensor for a point particle is
\begin{equation}
T^{\mu\nu} = m_p \int d\tau(-g)^{-1/2}u^{\mu}u^{\nu}
\delta^4(x-z(\tau)),
\end{equation}
here with $g = -r^4\sin^2\theta$ the determinant of the 
background Schwarzschild metric \eqref{eq:staticelement}. 
We now change coordinates $d\tau= (d\tau/dt)dt$, integrate 
over $t$, and use $u^t = dt/d\tau$, thereby finding
\begin{equation}\label{eq:IntTAB}
T^{\mu\nu} = m_p \frac{u^{\mu}u^{\nu}}{u^t r^2\sin\theta}
             \delta(r-r_p(t))
             \delta(\theta-\theta_p(t))
             \delta(\phi-\phi_p(t)).
\end{equation}
Combination of Eqs.~\eqref{eq:PA} and \eqref{eq:IntTAB}, with 
the assumption of equatorial motion, then gives
\begin{equation}
P^A = \frac{16\pi m_p}{\ell(\ell+1)}\frac{u^A u^\phi}{u^t}
       \bar{S}_{\phi}(\pi/2,\phi_p(t))\delta(r-r_p(t)),
\end{equation}
Because we have not integrated over $r$, the four--velocity 
components $u^\mu$ here may be viewed either as functions solely
of $r$, or solely of $t$ upon replacing $r$ by $r_p(t)$. Either
viewpoint will yield the same derivatives $\partial_B P^A$ insofar
as integration against test functions is concerned, and we view the 
components $u^\mu$ as depending on $r$. The delta functions of 
course depend on both $r$ and $t$. Having identified which terms 
depend on $r$ and $t$, we then calculate 
\begin{align}
\partial_r P^t & =  \frac{16\pi m_p}{\ell(\ell+1)}
                    (\partial_r u^{\phi})
                    \bar{S}_{\phi}\delta(r-r_p(t))
                   +\frac{16\pi m_p}{\ell(\ell+1)}
                    u^{\phi}\bar{S}_{\phi}
                    \partial_r \delta(r-r_p(t)) 
\\
\partial_t P^r & =  \frac{16\pi m_p}{\ell(\ell+1)}
       \frac{u^r u^\phi}{u^t}\left\{\left[\frac{u^\phi}{u^t}
       (\partial_\phi\bar{S}_{\phi}) - 
       \partial_r\left(\frac{u^r}{u^t}\right)
       \bar{S}_{\phi} \right]
       \delta(r-r_p(t))
       -\frac{u^r}{u^t}
       \bar{S}_{\phi}\partial_r\delta(r-r_p(t))
       \right\}.
\end{align}
To reach the last equation, we have replaced $\dot{\phi}_p(t)$
by $u^\phi/u^t$, which is permissible due to the presence of 
the accompanying delta function. Moreover, we have also made
the replacement
\begin{equation}
\dot{r}_p(t)\delta'(r-r_p(t)) \rightarrow
\frac{u^r}{u^t}\delta'(r-r_p(t)) + \left(\frac{u^r}{u^t}\right)'
\delta(r-r_p(t)),
\end{equation}
where the prime denotes partial $r$--differentiation. Finally, 
substitution of the last two results into Eq.~\eqref{eq:S_CPM},
along with Eq.~\eqref{eq:fourvelocity} and the identity
\begin{equation}
\partial_r \left(\frac{u^r}{u^t}\right)^2 
         = \frac{2}{E_p^2}
           \frac{f^2(r)}{r}
           \left[\left(1-\frac{5M}{r}\right)
                 \left(1+\frac{L_p^2}{r^2}\right) 
           -f(r)\right] + \frac{4M f(r)}{r^2},
\end{equation}
yields the desired results 
(\ref{eq:CPMsource},\ref{eq:CPMcoefficients}). We have also
used $S_{\phi\phi}(\pi/2,\phi)=\partial_{\phi}S_{\phi}(\pi/2,\phi)$, 
that is ordinary partial differentiation suffices in the equatorial plane.

%
%

\bibliographystyle{plain}

\begin{thebibliography}{99}

\bibitem{NASA_LISA}
Current website for the project: {\tt lisa.nasa.gov}, 
December 2008.

\bibitem{ScottHughes_LISA}
S.~A.~Hughes, {\em Lisa sources and science},
Proceedings of the 7\underline{th} Edoardo
Amaldi Conference on Gravitational Waves,
{\tt arXiv:0711.0188} (2007).

\bibitem{PoissonLRR}
E.~Poisson, {\em The Motion of Point
Particles in Curved Spacetime}, Living Rev. Relativity 7
(2004) 6 (136 pages). Available at
{\tt www.livingreviews.org/lrr-2004-6}.

\bibitem{ScottHughes_Adiabatic}
P.~A.~Sundararajan, G.~Khanna, and S.~A.~Hughes,
{\em Towards Adiabatic Waveforms for Inspiral
into Kerr Black Holes: A New Model of the Source
for the Time Domain Perturbation Equation},
Phys.~Rev.~D, 74 (2007) 104005 (20 pages).
Expanded version available as 
{\tt arXiv:gr-qc/0703028v3}.

\bibitem{Barack_sources}
L.~Barack and C.~Cutler,
{\em  LISA capture sources: Approximate
waveforms, signal--to--noise ratios, and
parameter estimation accuracy},
Phys.~Rev.~D, 69 (2004) 082005 (24 pages).

\bibitem{RW}
T.~Regge and J.~Wheeler, {\em Stability of a Schwarzschild Singularity},
Phys.~Rev.~108 (1957) 1063-1069.

\bibitem{ZER}
F.~J.~Zerilli, {\em  Effective Potential for Even-Parity Regge-Wheeler
Gravitational Perturbation Equations},
Phys.~Rev.~Lett., 24 (1970) 737-738.

\bibitem{Martel_CovariantPert}
K.~Martel and E.~Poisson, {\em
Gravitational perturbations of the Schwarzschild
spacetime: A practical covariant and gauge--invariant
formalism}, Phys.~Rev.~D, 71 (2005) 104003 (13 pages).
Expanded version available as {\tt arXiv:gr-qc/0502028}.

\bibitem{SopuertaLaguna}
C.~F.~Sopuerta and P.~Laguna,
{\em  Finite element computation of the gravitational
radiation emitted by a pointlike object orbiting a
nonrotating black hole},
Phys.~Rev.~D, 73 (2006) 044028 (17 pages).

\bibitem{MTW}
C.~W.~Misner, K.~S.~Thorne, and J.~A.~Wheeler, {\em Gravitation}
(Freeman, New York, 1973).

\bibitem{Lousto}
C.~O.~Lousto, {\em A time--domain fourth--order--convergent
numerical algorithm to integrate black hole perturbations
in the extreme--mass--ratio limit}, Class.~Quantum Grav.~22
(2005) S543-S568.

\bibitem{JungKhannaNagle}
J.--H.~Jung, G.~Khanna, and I.~Nagle, {\em A spectral collocation 
approximation of one--dimensional head--on collisions of black-holes},
to appear in Int.~J.~Mod.~Phys.~C, 
19 pages, {\tt arXiv:0711.2545} (2007).

\bibitem{CanizaresSopuerta}
P.~Ca\~{n}izares and C.~F.~Sopuerta, {\em Simulations of 
Extreme--Mass--Ratio Inspirals Using Pseudospectral Methods}, 
J.~Phys.~Conf.~Series 154 (2009) 012053 (6 pages), {\tt 
arXiv:0811.0294}.

\bibitem{HesthavenWarburtonEM}
J.~S. Hesthaven and T.~Warburton, {\em 
  High-Order Accurate Methods for Time-domain
  Electromagnetics}, Comp.~Mod.~Engin.~Sci.~5 (2004) 395-408.

\bibitem{HEST}
J.~S. Hesthaven and T.~Warburton,
{\em Nodal Discontinuous Galerkin Methods:
Algorithms, Analysis, and Applications}
(Springer-Verlag, New York, 2008).

\bibitem{Zumbusch} G.~Zumbusch,
{\em Finite Element, Discontinuous Galerkin, and Finite 
Difference evolution schemes in spacetime}, 15 pages,
{\tt arXiv:0901.0851} (2009).

\bibitem{Chan_MathBlack}
S.~Chandrasekhar, {\em The Mathematical Theory 
of Black Holes} (Oxford University Press, Oxford, 2000).

\bibitem{Martel_GravWave}
K.~Martel, {\em  Gravitational waveforms
from a point particle orbiting a Schwarzschild
black hole}, Phys.~Rev.~D, 69 (2004) 044025 (20 pages).

\bibitem{CUT}
C.~Cutler, D.~Kennefick, and E.~Poisson,
{\em Gravitational radiation reaction for
bound motion around a Schwarzschild black hole},
Phys.~Rev.~D, 50 (1994), 3816-3835.

\bibitem{RK}
M.~H.~Carpenter and C.~Kennedy, {\em
Fourth Order 2$N$--Storage Runge--Kutta Schemes},
NASA Technical Memorandum 109112 (1994).

\bibitem{FanGDG}
K.~Fan, W.~Cai, X.~Ji, {\em A generalized
discontinuous Galerkin (GDG) method for
Schr\"{o}dinger equations with nonsmooth solutions},
J.~Comp.~Phys., 227 (2008) 2387-2410.

\bibitem{LAU}
S.~R.~Lau,
{\em Analytic structure of radiation boundary kernels for
blackhole perturbations}, J.~Math.~Phys., 46 (2005)
102503 (21 pages).

\bibitem{AGH}
B.~Alpert, L.~Greengard, and T.~Hagstrom,
{\em Rapid Evaluation of Nonreflecting
Boundary Kernels for Time-Domain Wave
Propagation}, SIAM J.~Numer.~Anal., 37
(2000) 1138-1164.

\bibitem{Poisson1993} E.~Poisson, {\em Gravitational radiation from
a particle in circular orbit around a black hole. I. Analytical 
results for the nonrotating case}, Phys.~Rev.~D, 47 
(1993) 1497-1510.

\bibitem{DRPP1971} M.~Davis, R.~Ruffini, W.~H.~Press, and
R.~H.~Price, {\em Gravitational Radiation from a Particle Falling 
Radially into a Schwarzschild Black Hole}, Phys.~Rev.~Lett.~27 
(1971) 1466-1469.

\bibitem{AbrahamsEvans1988}
A.~M. Abrahams and C.~R. Evans, {\em
Reading off gravitational radiation waveforms in numerical
relativity calculations: Matching to linearized gravity},
Phys.~Rev.~D, 37 (1988) 318-332.

\bibitem{AbrahamsEvans1990}
A.~M. Abrahams and C.~R. Evans, {\em
Gauge-invariant treatment of gravitational
radiation near the source: Analysis and numerical simulations},
Phys.~Rev.~D, 42 (1990) 2585-2594.

\bibitem{SHIBATA}
T.~Tanaka, M.~Shibata, M.~Sasaki, H.~Tagoshi,
and T.~Nakamura, {\em Gravitational Wave Induced by
a Particle Orbiting around a Schwarzschild Black Hole},
Prog.~Theo.~Phys.~90 (1993) 65-83.

\bibitem{BMNOS2002} L.~Barack, Y.~Mino, H.~Nakano, A.~Ori, and
M.~Sasaki, {\em Calculating the Gravitational Self--Force in
Schwarzschild Spacetime}, Phys.~Rev.~Lett.~88 (2002)
091101 (4 pages).

\bibitem{BarackSago}
%
%
L.~Barack and N.~Sago, {\em Gravitational self-force on a
particle in circular orbit around a Schwarzschild black hole},
Phys.~Rev.~D, 75 (2007) 064021 (25 pages).

\bibitem{Detweiler}
%
%
S.~Detweiler, {\em Consequence of the gravitational self-force
for circular orbits of the Schwarzschild geometry},
Phys.~Rev.~D, 77 (2008) 124026 (15 pages).

\bibitem{NDT2007}
A.~Nagar, T.~Damour, and A.~Tartaglia, {\em Binary black hole
merger in the extreme mass ratio limit}, Class.~Quant.~Grav.~24
(2007) S109-S124.

\bibitem{BarackDetComparison}
%
%
N.~Sago, L.~Barack, S.~Detweiler, {\em Two approaches for the
gravitational self-force in black hole spacetime: Comparison of
numerical results}, Phys.~Rev.~D, 78 (2008) 124024 (9 pages).

\bibitem{ThorneMultipole}
K.~S. Thorne, {\em Multipole expansions of gravitational radiation},
Rev.~Mod.~Phys.~52 (1980) 299-339.

%

\end{thebibliography}
\end{document}